\documentstyle[12pt]{article}
\let\LARGE=\Large
\let\Large=\large
\let\large=\normalsize
\newcommand{\be}[3]{\begin{equation}  \label{#1#2#3}}     
\newcommand{\ee}{ \end{equation}}
\newcommand{\ba}{\begin{array}}
\newcommand{\ea}{\end{array}}

\def\beq{\begin{equation}}
\def\eeq{\end{equation}}
\def\beqa{\begin{eqnarray}}
\def\eeqa{\end{eqnarray}}

\renewcommand{\d}{\delta}

\newcommand{\m}{\mu}

\newcommand{\n}{\nu}

\begin{document}
\thispagestyle{empty}
\rightline{HUB-EP-98/18}
\vspace{1mm}
\rightline{hep-th/9803072}
\vspace{2truecm}
\centerline{\bf \LARGE
String Vacua with $N=2$ Supersymmetry in Four Dimensions}

\vspace{1.2truecm}

\centerline{\bf
Dieter L\"ust\footnote{luest@qft1.physik.hu-berlin.de}}

\vspace{.5truecm}
{\em 
\centerline{Humboldt-Universit\"at, Institut f\"ur Physik,
D-10115 Berlin, Germany}}

\vspace{1.2truecm}
\vspace{.5truecm}
\begin{abstract}
In this article we review 
four-dimensional string vacua with 
$N=2$ space-time
supersymmetry. In particular,
we will discuss several aspects of the string-string duality between
the heterotic string, compactified on $K3\times T^2$, 
and the type II superstring
compactified on a Calabi-Yau three-fold. We investigate the massless 
supersymmetric spectra, showing agreement for a 
large class of dual heterotic/type
II string pairs.  Some emphasis is given to
non-perturbative heterotic phenomena, such as non-perturbative
transitions among different vacua and strong
coupling singularities, and to their
geometric Calabi-Yau description on the type II side.
We compare the effective $N=2$ supergravity actions
of dual heterotic/type II string compactifications, and show that the
$N=2$ prepotentials and also higher order gravitational couplings
nicely agree in the weak heterotic coupling limit.
Finally  
we consider extremal black hole solutions of $N=2$
supergravity which arise in the context of 
heterotic or type II $N=2$
compactifications. 
For the type II backgrounds we show how
the entropies of these black holes depend on the
topological data of the 
underlying Calabi-Yau spaces; we also construct massless
black holes which are relevant for the conifold transition among
different Calabi-Yau vacua.

\vskip0.5cm
\noindent {\it The work is based on material presented at the conference
`SUSY '97' in Philadelphia (27-31 May 1997), at the `31st International
Symposium Ahrenshoop' in Buckow (27 -31 August 1997) and at the 
TMR-meeting on `Quantum Aspects of Gauge
Theories, Supersymmetry and Unification' in Neuchatel (18-23 September 1997).}
  
\end{abstract}

\bigskip \bigskip
\newpage

\tableofcontents
\newpage

\section{Introduction}

Several types of strong-weak coupling duality symmetries in string theories
were explored and established during recent years
(for recent review see \cite{refdualities}). For example,
the heterotic string compactified to four dimensions on a six-dimensional
torus $T^6$ exhibits $S$-duality \cite{sduality}, which is the stringy
version of the Montonen-Olive electric/magnetic duality symmetry in $N=4$
supersymmetric field theories. 
$S$-duality transforms perturbative, electrically charged states into
non-perturbative solitonic states with non-vanishing magnetic charges.
To give a slightly more general definition,
$S$-duality is a non-perturbative symmetry which relates the weak coupling
regime of one particular theory to the strong coupling regime of
the same theory:
$g\leftrightarrow {1\over g}$; 
hence $S$-duality can be regarded as a self-duality
within in a particular string model.
In this spirit, also the uncompactified IIB superstring possesses
$S$-duality \cite{jschwarz}, where  $S$-duality transforms the 
BPS p-branes in the Neveu-Schwarz (NS)
sector into the Dirichlet (D) p-branes of the Ramond (R) sector (for  recent
introductions into D branes see for example \cite{thor}).

Second, there exist also string duality symmetries which relate
one string construction, say at weak coupling, to a seemingly
different, but eventually equivalent string construction at strong
coupling. The first
and very  basic example of this string-string duality symmetry is
the duality between the heterotic string, compactified to six dimensions
on $T^4$, and the type IIA superstring, compactified on $K3$ 
\cite{DuffKhuri,HullTown,witt}.
To establish the string-string duality symmetry one has to take into account
the full perturbative and non-perturbative BPS spectra of the dual models.
In general, the perturbative states of one string construction are mapped
to non-perturbative states of the dual string construction, and vice versa,
by the string-string duality symmetry.
In fact one can show that the type IIA string contains the heterotic string
as a soliton \cite{HarStro}, and one obtains the following duality relation
among the heterotic and type IIA string couplings in six dimensions:
$g_{H}={1\over g_{IIA}}$.

Finally, there are duality symmetries which go beyond the 
perturbatively known string
constructions. Namely, Witten \cite{witt} has given very strong evidence
that    the
type IIA string in ten dimensions is dual to 11-dimensional supergravity
compactified on a circle of radius $R_{11}$, where the Kaluza-Klein modes
of 11-dimensional supergravity correspond to the BPS Dirichlet 0-branes
of the type IIA string. The full quantum version of 11-dimensional
supergravity is now called $M$-theory; $M$-theory on circle of radius $R$
is supposed to 
describe the full strongly coupled type IIA superstring, with the
following duality relation among the IIA string coupling and $R_{11}$:
$R_{11}\sim g_{IIA}^{2/3}$. In addition, one can also show \cite{horavawitten}
that $M$-theory compactified on the semicircle $S^1/{\bf Z}_2$ corresponds to
the strongly coupled heterotic string with gauge group $E_8\times E_8$.
Including also all the other string theories,
it seems nowadays clear that all five known
consistent string constructions, the two heterotic strings, the type IIA,B
superstrings and the type I superstring, plus 11-dimensional supergravity
are related by some kind of duality transformation such that they are all
equivalent and on equal footing. The underlying unifying quantum
theory is 
$M$-theory, which contains all the six theories as different ways to perform a
weak coupling limit. However a fundamental definition of $M$-theory
is up to now  not clear. One promising attempt to define $M$-theory
is given by a Hamiltonian theory of non-commuting matrices, known as
M(atrix)-theory \cite{matrix}. 
In particular these matrices describe the dynamics
of the non-perturbative D 0-branes of the type IIA superstring.
It is important to note that in M(atrix)-theory space-time is not fundamental,
but arises only in the limit where the matrices are commuting.

In order to fully establish the unity of all string vacua one must also
show that
there exist continuous transitions among different string compactifications,
i.e. string backgrounds. 
For example, in the study of the moduli spaces of type II
strings compactified on Calabi-Yau three-folds it became clear that the
topologies of the Calabi-Yau spaces can be continuously deformed into
each other; therefore many, or even possibly all, Calabi-Yau vacua are
just branches of a larger universal moduli space.
Calabi-Yau phase
transitions contain for example the conifold
transition in the complex structure moduli space of type IIB
superstrings. The conifold transitions
occur at those points in the moduli space where
certain 3-cycles shrink to zero size and then are blown up as two
cycles, changing in this way the Hodge numbers.   
The physical understanding of the
conifold transition was provided by Strominger \cite{strominger}; at
the conifold point a BPS hypermultiplet black hole becomes massless,
being responsible for the singularity in the moduli space metric of
the $N=2$ vector multiplets at this point.

In this paper we will review  the string-string duality between
dual string pairs in four dimensions with $N=2$  space-time
supersymmetry. In particular we will concentrate  on
the duality between certain $N=2$ string vacua \cite{KV,FHSV}, 
namely the heterotic string
compactified on $K3\times T^2$, which is dual to the type IIA string
on a particular Calabi-Yau background,
as it was first discussed by Kachru and Vafa \cite{KV}.\footnote{In addition, 
four-dimensional string
models with $N=2$ space-time are given by type I compactifications
on $K3\times T^2$; this leads then to a $N=2$ heterotic/type II/type I
string triality \cite{triality}.} Using this $N=2$ string-string duality
symmetry, many interesting non-perturbative phenomena can be studied. 
In particular, effects which are
non-perturbative on the heterotic side, follow from
classical geometrical considerations on the type II side,
where various branes are wrapped around the internal
cycles of the underlying Calabi-Yau space. In this way, performing a
suitable field theory limit, this 
approach provides a nice geometrical
understanding \cite{klmvw,witt1} of non-perturbative
effects in supersymmetric field theories, 
for example ala Seiberg/Witten \cite{sw} (for  reviews on non-perturbative
effects in $N=2$ supersymmetric gauge theories and their string origin
see \cite{n2sw}).

Obviously, for a candidate dual string pair the massless spectrum has to 
match. In addition, more quantitative checks are possible by comparing
the effective interactions of the massless modes. Specifically
the holomorphic couplings of the $N=2$ vector multiplets are determined by
the $N=2$ prepotential of $N=2$ special geometry
(for a review on the holomorphic couplings in
$N=2$ string vacua see \cite{curio}). We will 
in section 3 show that the
heterotic and the type IIA prepotentials agree \cite{CCLM,CCL} for several
dual pairs in a certain corner of the vector multiplet moduli space,
which corresponds to the heterotic weak coupling limit. Moreover explicit
comparisons \cite{KLT,AGNT,C,CCLMR,CCLM,CCL,DCLMR} of higher 
derivative vector-gravitational couplings provide an additional
successful confirmation of the $N=2$ string-string duality.

In section 4 we will study the extremal BPS saturated solutions 
\cite{FeKaStr,n2solut,BCDWKLM,BehrndtLuestSabra,BehLu,msw,higher}
of
the four-dimensional $N$=2 supergravity 
coupled to $N=2$ vector multiplets as the effective langrangian of the $N=2$
string models. In general one obtains
extremal, supersymmetric,
charged black holes that allow, 
besides the non-trivial four-dimensional black hole metric,
for non-constant moduli fields.
We will 
determine the space-time
dependence of the black hole metrics, of the  scalar
fields and of the gauge fields 
and also the macroscopic
Bekenstein-Hawking black hole entropies from the  prepotential of $N=2$
special geometry. 
In case of type IIA Calabi-Yau compactifications,
the black hole entropies therefore depend on the topological data 
of the underlying Calabi-Yau spaces, such as intersection numbers
and rational instanton numbers. In the dual heterotic vacua, these
contributions correspond to perturbative as well non-perturbative
corrections to the black hole entropies. 
The $N=2$ black hole solutions are also  relevant for transitions 
between different type II vacua on Calabi-Yau spaces in
four dimensions. We will  discuss the type IIB conifold
transition where one of the complex structure moduli fields is small.
The following generic picture will emerge. If one moves through a non-singular
space time one varies at the same time the radii of the cycles of the
Calabi-Yau. At any point in space time the Calabi-Yau looks
differently.  We will find that for vanishing 3-cycles in IIB
compactifications, our
solution corresponds to a massless BPS states.

\section{$N=2$ Heterotic/Type II String Duality}

In this chapter we will compare the spectra of $E_8\times E_8$
heterotic string compactifications on $K3\times T^2$ with the dual type IIA
compactifications on a suitably chosen Calabi-Yau 3-fold. 
On the heterotic
side, the different vacua are essentially characterized by different
choices of the $E_8\times E_8$ gauge bundle. These choices lead to a large 
variety
of different spectra in four dimensions. On the type II side, the different
ways to compactify are defined by the choice of the Calabi-Yau 3-fold $X^3$.
We will discuss how the heterotic data of specifying the gauge bundle
will map on the Calabi-Yau data of the dual type II model. 
It turns out that the Calabi-Yau spaces that are dual to perturbative
heterotic string vacua are $K3$-fibrations over a two sphere $P^1$ 
\cite{klm,asplou}.

Since the toroidal $T^2$ compactification on the heterotic side leads
to a universal sector, the heterotic data can be already specified by
considering the heterotic compactification on $K3$ to six dimensions.
Many of the interesting perturbative and non-perturbative effects like
gauge symmetry enhancement and transitions can be already understood
in six dimensions.
On the type II side one can also construct the corresponding dual models
in six dimensions by considering $F$-theory compactified on the same
Calabi-Yau space $X^3$ which has to be an elliptic fibration over an
two-dimensional base $B^2$. 

The $N=2$ heterotic/type II string duality can be explained, at least in an
heuristic way, from the $N=4$ string -string duality between the
heterotic string on $T^6$ and the type IIA string on 
$K3\times T^2$. This more simple string
duality already contains the main clues of how the non-Abelian
gauge symmetry enhancement is realized on the type II side.
Hence, to give an understanding of this important phenomenon and to
see the how the $N=2$ string duality is related to the $N=4$
string duality, let us briefly compare the heterotic string on $T^6$
with type IIA on $K3\times T^2$.

\subsection{The $N=4$ heterotic/type II string duality}

The heterotic string on $T^6$ contains as its massless spectrum first the
$N=4$ supergravity multiplet, which includes six graviphotons plus
the complex dilaton axion field, $S_H=e^{-\phi_H}+ia$, and second 22 $N=4$
vector multiplet with 22 $U(1)$ vector boson and 132 moduli fields. Among
the 132 moduli there is the complex $T_H$ field whose real part corresponds to
the K\"ahler class of a $T^2$, which describes, say, the
compactification from six to four dimensions. The well known Narain moduli
space, including the $S_H$-field, is locally given by \cite{narain}
\begin{equation}
{\cal M}_{N=4}={SO(6,22)\over SO(6)\times SO(22)}\otimes
\Biggl({SU(1,1)\over U(1)}\Biggr)_{S_H}.\label{classmod}
\end{equation}
Due to the presence of the 
duality transformations ${\cal M}_{N=4}$ has to be modded by
the discrete duality group $\Gamma=SO(2,22;{\bf Z})_T\times SL(2,{\bf Z})_S$,
where the first factor corresponds to the perturbative $T$-duality group,
and the second factor is the non-perturbative $S$-duality group.
The gauge group, at generic points in the moduli space, is given
by $G=U(1)^{22}\times U(1)^6$, where the last factor belongs to the six
graviphotons. At special points in the moduli space, which correspond to
fix-points of the $T$-duality group, there is a perturbative 
gauge symmetry enhancement to a non-Abelian gauge group of maximal rank 22.
The non-Abelian charged gauge bosons are given by internal momentum and
winding states; these states hence belong to short perturbative BPS multiplets
of the $N=4$ supersymmetry algebra with central charges.

The type IIA string compactified on $K3\times T^2$ precisely leads to same
massless spectrum with the same
moduli space as the heterotic string discussed before \cite{seiberg}.
However now, in contrast to the heterotic case, the $N=4$ gravity
multiplet contains the $T_{IIA}$-modulus of two-torus $T^2$  \cite{flt}. 
On the other hand, the complex type IIA dilaton field $S_{IIA}$
sits in one of the 22 vector multiplets, where the 132 scalars divide 
themselves
into 84 Neveu-Schwarz scalars and 48 Ramond scalar fields.
This means that perturbative effects in the heterotic string can be 
non-perturbative in the type II string, and vice versa.
This exchange between the $S$ and $T$ fields implies
that the string-string duality relation between the heterotic
and type couplings reads: 
\begin{equation}
S_{H(IIA)}=T_{IIA(H)}.\label{dualrel4}
\end{equation}
This relation can be easily understood from the 
strong-weak coupling string-string duality relation
in six-dimensions, $\phi_H^6=-\phi_{IIA}^6$, and by noting that the
four-dimensional dilaton  and the six-dimensional dilaton are related as
\begin{equation}
{\rm Re}S/{\rm Re}T=e^{-\phi^6}.\label{dilaton64}
\end{equation}
It is also instructive to recognize that the subspace of ${\cal M}_{N=4}$
which is spanned by the 84 NS scalar fields is given by the coset
${SO(4,20)\over SO(4)\times SO(20)}\otimes {SO(2,2)\over SO(2)\times SO(2)}$.
This is just the moduli space of $K3\times T^2$.

Let us now come to the type IIA gauge symmetries. The perturbative
gauge group is always
$U(1)^{22}\times U(1)^6$; four graviphotons are from the NS-NS sector and all
remaining $U(1)$ vector fields come from the R-R sector. Specifically the
additional 24 R-R vector fields arise from the ten-dimensional
R-R gauge fields $A_M$ and $A_{MNP}$. One of them is the four-dimensional
component of $A_M$, and the remaining 23 come from expressing $A_{MNP}$ as
exterior product of a four-dimensional one-form gauge potential
times each of the 22+1 harmonic two-forms of $K_3\times T^2$.
To discuss the non-Abelian gauge symmetry enhancement in the type IIA string,
we have to find the charged BPS states which couple to 22 Abelian R-R
gauge bosons \cite{HullTown}. 
These cannot be found within the perturbative 
type II spectrum, but they are 
given by the non-perturbative electric D 2-branes which can be wrapped
around the 22 homology two-cycles of $K3$.  They lead in four dimensions
to electrically charged BPS black holes with 22 different electric charges.
The masses of the BPS particles is proportional to the area of the $K3$
two-cycles. Therefore  massless non-Abelian gauge bosons arise at those
points in the $K3$ moduli space where the $K3$ degenerates, and 
the two-cycles shrink to zero sizes. This is the non-perturbative version
of the perturbative Frenkel-Kac mechanism in conformal field theory.
In fact, the degenerate $K3$'s allow for an ADE classification, and the
intersection form in the cohomology of $K3$ is just given by
$E_8\oplus E_8\oplus H\oplus H\oplus H$ ($H$ being the hyperbolic plane).
This same observation also led to the discovery of the heterotic string
as a soliton of the type IIA string in six dimensions \cite{HarStro}.

The $N=2$ string duality between the heterotic string on
$K3\times T^2$ and the type IIA string on $X^3$ can be made plausible
from the $N=4$ string duality by the
adiabatic extension principle \cite{vawi}.
Namely, starting from the $N=4$ duality, replace the common $T^2$ factor
by the two sphere $P^1$, and let the rest of the space vary as a fibre
over the base $P^1$. This is the fibre wise application of the
$N=4$ string duality. On the type IIA side, the resulting space is
a $K3$ fibred Calabi-Yau space $X^3$ with base $P^1$; the various ways
to perform this $K3$-fibration 
results in the variety of different type II $N=2$
vacua. On the heterotic side, we deal with a $T^2$ varying over the 
base $P^1$;
this is nothing else than $K3$ represented as an elliptic fibration.
The different heterotic
models arise by specifying how the fibration, which breaks
$N=4$ supersymmetry to $N=2$, acts on the heterotic gauge bundle.
One has to emphasize that this adiabatic  argument
is somewhat heuristic; nevertheless one can show that for every perturbative
$N=2$ heterotic vacuum on $K3\times T^2$ the dual Calabi-Yau space has
to be a $K3$ fibration, as we will discuss in the following sections.

\subsection{The heterotic string on $K3\times T^2$}


\subsubsection{Spectrum in Six Dimensions}


Now let us discuss in more detail the spectrum of the $E_8\times
E_8$ heterotic string
compactified on $K3\times T^2$. We start with the compactification on $K3$
which leads to (0,1) supergravity in six dimensions. In general,
the 6-dimensional massless spectrum contains first the gravity supermultiplet
with $g_{\mu\nu}$ and $B_{\mu\nu}^-$ as bosonic components.
The massless matter fields are given by $N_V^6$ vector multiplets each with
a six-dimensional vector field, by $N_H$ hypermultiplets containing each
four real scalar fields, and by $N_T$ tensor multiplets consisting each of
a real scalar plus a self-dual antisymmetric tensor $B_{\mu\nu}^+$.
Since the effective field theory is a chiral 
supergravity there are strong constraints due to anomaly matching
conditions. 
The anomaly 8-form is given by the expression
\begin{equation}
I_8=\alpha {\rm tr}R^4+\beta ({\rm tr}R^2)^2+\gamma{\rm tr}R^2{\rm tr}F^2
+\delta ({\rm tr}F^2)^2.\label{gravan}
\end{equation}
The requirement of absence of gravitational anomalies, i.e. $\alpha=0$,
demands that
\begin{equation}
N_H-N_V^6+29 N_T=273.\label{gravmatch}
\end{equation}
Then the anomaly 8-form factorizes, $I_8=I_4\wedge\tilde I_4$,
$I_4={\rm tr}R^2-\sum_av_a{\rm tr}F_a^2$, $\tilde I_4={\rm tr}R^2-\sum_a
\tilde v_a{\rm tr}F_a^2$, where the coefficient $v_a$, $\tilde v_a$ are
gauge group dependent constants \cite{erler}, and the remaining
 anomalies
 can by cancelled by the Green-Schwarz mechanism \cite{greenschw}. 
 For that purpose the three
form $H=dB+\omega^{L}-\sum_av_a\omega^{YM}_a$ must be well defined, 
i.e. $\int_{K3}H=0$, and
the following condition must hold:
\begin{equation}
\sum_an_a=\sum_a\int({\rm tr} F^2_a)=\int_{K3}{\rm tr} R^2=24.\label{gsmatch}
\end{equation}
Here $n_a$ $(a=1,2)$ are the numbers of $E_8\times E_8$ instantons turned on.
So we see that the gauge bundle has to be non-trivial, and the gauge group
$E_8\times E_8$ will be broken to some subgroup $G^1_{n_1}\times G^2_{n_2}$
by the gauge instantons.
This unbroken group depends on the distribution of the instanton numbers
$n_1$, $n_2$, and also on the values of hypermultiplet moduli, as we will
discuss later.
The number $N_V^6$ is then simply given by the dimension of this 
unbroken group.

In addition one can consider also the 
non-perturbative background of $n_5$ heterotic 5-branes
turned on \cite{seiwi}. In this case
the $E_8\times E_8$ heterotic string spectrum on $K3$ is determined
by the three integers $n_1$, $n_2$ and $n_5$, and the anomaly condition
eq.(\ref{gsmatch}) has to modified in the following way \cite{sagnotti}:
\begin{equation}
n_1+n_2+n_5=24.\label{gsmatch5}
\end{equation}

Without 5-branes, i.e. for perturbative vacua with $n_5=0$, 
the number of tensor multiplets
is just  one, where the corresponding scalar field is   the heterotic
dilaton $\phi^6_H$. However considering the non-perturbative contribution
of heterotic 5-branes, there are
  additional tensor multiplets in
the massless spectrum \cite{seiwi}, 
since on the world sheet theory of the 5-brane
lives a massless tensor field. Hence 
\begin{equation}
N_T=1+n_5.\label{nt}
\end{equation}
Thus for non-trivial 5-brane backgrounds, the Coulomb branch of the
six-dimensional theory is characterized by 
a real
$(1+n_5)$-dimensional moduli space, parametrized by the scalar field vev's
of the tensor multiplets.
The Higgs branch of the theory is parametrized by the scalar fields
of the hypermultiplets. 
The moduli space of the hypermultiplets is a quaternionic space, and its 
quaternionic dimension, i.e the number of hypermultiplets is given by
\begin{equation}
N_H=20+n_5+\dim_Q{\cal M}^{\rm inst}_{n_1}\lbrack H_1\rbrack 
+\dim_Q
{\cal M}^{\rm inst}_{n_2}\lbrack H_2\rbrack ;\label{nh}
\end{equation}
in this formula, the first term comes from the 20 moduli of $K3$.
The second term arises, since the position of each five brane on $K3$
is parametrized by a hypermultiplet. The last two terms denote the
dimensions of the quaternionic instanton moduli space of the embedded
$E_8$ instantons. Specifically, if $E_8$ is broken to the group $G$
by $n$ instantons, the instanton moduli space has the dimension
\begin{equation}
\dim_Q{\cal M}^{\rm inst}_n\lbrack H \rbrack =nc_2(H)-\dim H,\label{diminst}
\end{equation}
where $H$ is the commutant of $G$ in $E_8$, and $c_2(H)$ is the quadratic
Casimir of $H$.

Let us discuss the unbroken gauge group in little bit more detail. 
For $n\geq 10$, $E_8$ is completely broken at an arbitrary point in the
hypermultiplet moduli space. However at special loci in the 
hypermultiplet moduli space the instantons fit into a subgroup of $E_8$,
such that there is an unbroken gauge group left over. This effect
is nothing else than the (reverse) Higgs effect in field theory, i.e.
breaking the gauge group by the vev's of the charged hypermultiplets. 
For example, if the instantons fit into a $E_7$ subgroup,
a $SU(2)$ gauge symmetry gets restored; the number of hypermultiplets
necessary to be tuned to open a $SU(2)$ can be easily read off from 
eq.(\ref{diminst}):
\begin{equation}
\Delta N_H(SU(2)) =  \dim_Q{\cal M}^{\rm inst}_n\lbrack E_8\rbrack
 - \dim_Q{\cal M}^{\rm inst}_n\lbrack E_7\rbrack 
  = 12 n -115.\label{deltanh}
\end{equation}
On the other hand, for $n<10$ there are not enough instantons to break
$E_8$ completely; so in this case there is a terminal gauge group which
is given by $G=E_8,E_7,E_6,SO(8)$ for $n=0,4,6,8$. (No smooth $E_8$
instantons exist for $n=1,2,3$.)
Again at special loci in the hypermultiplet moduli these gauge groups can
be further enhanced.

As already mentioned the gauge group breaking by the $E_8$ instantons
can be equally described by the standard Higgs effect. Here the starting point
is to consider $SU(2)$ gauge bundles on $K3$ with instanton numbers 
$(n_1,n_2)$.
These break $E_8\times E_8$ to $E_7\times E_7$ with the following
hypermultiplet spectrum (in addition there are 20 singlets, being
the $K3$ moduli, plus $n_5$ singlets from the 5-branes):
\begin{equation}
  {1\over 2}(n_1-4)({\bf 56},{\bf 1})+{1\over 2}(n_2-4)({\bf 1},{\bf 56})
+
(2(n_1+n_2)-6)({\bf 1},{\bf 1}).\label{spectrume7}
\end{equation}
The $E_7\times E_7$ gauge group can now be broken by giving vev's to the
charged hyper multiplets. For $n\geq 10$, $E_7$ can be completely Higgsed
through the chain $E_7\rightarrow E_6\rightarrow SO(10)\rightarrow SU(5)
\rightarrow SU(4)\rightarrow SU(3)\rightarrow SU(2)\rightarrow 1$.
The dimensions of the corresponding Higgs moduli spaces precisely agree with
the dimensions of the $E_8$ instanton moduli spaces, given in 
eq.(\ref{diminst}).
On the other hand, for $n<10$, there are not enough charged hypermultiplets
for a complete gauge symmetry breakdown, and one ends at the already quoted
terminal gauge groups.
Let us summarize the six-dimensional heterotic
spectrum in the following small table for the case of no
five branes, i.e. $n_5=0$, $n_1=12+k$, $n_2=12-k$, assuming that the gauge
groups are Higgsed as far as possible:

\vskip0.2cm
\begin{center}
\begin{tabular}{|c||c|c|} \hline
  $k$     & $G_2$  & $N_H$ \\ \hline\hline
0,1,2  &   1 & 244   \\ \hline
4 &    $SO(8)$& 272 \\ \hline
6 & $E_6$ & 322 \\ \hline
8 & $E_7$ & 377 \\ \hline
12 & $E_8$ & 492 \\ \hline
\end{tabular}
\end{center}

\vskip0.1cm
\noindent {\bf Table 1:} Perturbative heterotic spectra in 6 dimensions.

\vskip0.2cm
\noindent
As expected, all spectra are in agreement with gravitational anomaly
matching condition eq.(\ref{gravmatch}).


\subsubsection{Spectrum in Four Dimensions}


Upon compactification from six to four dimensions on $T^2$, 
the massless heterotic spectrum can be obtained in a straightforward way.
Beside the massless $N=2$ supergravity multiplet, including one $U(1)$
graviphoton field, there are $N_V$ massless vector and $N_H$ massless
hypermultiplets. The structure and the number of the hypermultilpets
is unchanged in comparison to six-dimensions; the hypermultiplets parametrize
the quaternionic moduli space ${\cal M}_H$.
On the other hand, a four-dimensional $N=2$ vector multiplet contains
in contrast to six dimensions, a complex scalar field. Therefore the
massless vector multiplets give rise to a new complex (special) K\"ahler
moduli 
space ${\cal M}_V$ of complex dimension $N_V$,
parametrizing the Coulomb branch
of the theory. Giving vevs to the complex scalar fields the non-Abelian
gauge group is broken to the maximal Abelian subgroup $U(1)^{{\rm rank }G}$.
So let us determine the number, $N_V$, of four-dimensional Abelian
vector fields from the six-dimensional spectrum. First the six-dimensional
unbroken gauge group $G_1\times G_2$ provides ${\rm rank}(G_1\times G_2)$
vector multiplets; 
the corresponding $U(1)$ (Wilson line) moduli are
denoted by $V_i$. Second from the $T^2$ compactification one obtains
two more $U(1)$ vector multiplets, called $T$ and $U$, whose complex scalar
components describe the shape of the internal $T^2$. Finally each tensor
multiplet in six-dimenensions leads to  a $U(1)$ vector field. The one
which contains the dilaton $\phi^4$ is commonly denoted by $S$.
Additional  $U(1)$ vectors from six-dimensional tensor fields are of
non-perturbative nature.
So in total we obtain that
\begin{equation}
N_V={\rm rank}(G_1\times G_2)+2+n_5.\label{nv}
\end{equation}

The total moduli space in four dimension is given by the direct product
of ${\cal M}_V$ and ${\cal M}_H$, ${\cal M}={\cal M}_V\otimes {\cal M}_H$,
because due to $N=2$ supersymmetry the kinetic terms do not mix hyper
with vector multiplets. For perturbative vacua with no non-perturbative
vector fields,
the classical vector
moduli space can be simply obtained by 
truncating the $N=4$ Narain moduli space
to its $N=2$ subsector:
\begin{equation}
{\cal M}_V={SO(2,N_V-1)\over SO(2)\times SO(N_V-1)}\otimes
\Biggl({SU(1,1)\over SU(1)}\Biggr)_S,\label{mv}
\end{equation}
where the relevant target space duality group has still
to be modded out.
Along the Coulomb branch,  at special points in ${\cal M}_V$
the Abelian gauge group $U(1)^{N_V}$ is classically
enhanced to a non-Abelian gauge
group, where the rank is always preserved. 
Consider the simplest case, namely the perturbative vacua with 
the three vector fields $S$, $T$ and $U$, i.e. $N_V=3$,
which originate from the six-dimensional models with $n_5=0$ and 
completely broken gauge groups ($k=0,1,2$).
The perturbative target space duality group which acts on the moduli $T$ and
$U$ is given by $\Gamma=SO(2,2;{\bf Z})=SL(2,{\bf Z})_T\times SL(2,{\bf Z})_U
\times {\bf Z}_2^{T\leftrightarrow U}$.
At the fixed point/lines of $\Gamma$ the $U(1)^2$ gauge symmetry which belongs
to $T$ and $U$ gets enhanced. The charged  states which can become
massless at the critical points/lines are given by elementary BPS vector
multiplets,
whose BPS masses are equal to the central charge of the $N=2$
supersymmetric algebra.
 One finds
that at the line $T=U$ the enhanced gauge
group is
$SU(2)\times U(1)$, for $T=U=1$ there is a
further enhancement to $SU(2)^2$, and at the point $T=U=e^{i\pi/6}$ the
enhanced group is $SU(3)$. Also note that particular elements
of the duality group $\Gamma$, namely
the fixed point transformations, correspond to the elements of the Weyl group
of the enhanced gauge groups. For example the exchange $T\leftrightarrow
U$ is just the Weyl reflection of the $SU(2)$ which appears at the line
$T=U$.

It is important to emphasize that the perturbative gauge boson enhancement
on special loci of ${\cal M}_V$ in general disappears when discussing all
non-perturbative corrections to the moduli space ${\cal M}_V$. 
Non-perturbatively,
according to Seiberg/Witten \cite{sw}, the points of enhanced gauge symmetries
will split, and there will be special loci where BPS hyper multiplets, namely
BPS monopoles and dyons, become massless.
By giving vevs to these massless hyper multiplets (monopole
condensation)  non-perturbative transitions to a different vacuum can 
take place under certain circumstances (when the gauge theory also possesses
charged hypermultiplets); during this transition
$N_V$ is reduced, but $N_H$ is increased.
In the dual type II models 
the loci of massless monoples and dyon
are given by the conifold loci of
the dual Calabi-Yau spaces, and the possible transitions 
correspond to conifold
transitions where the topology of the underlying Calabi-Yau space gets
changed.

Now we turn to the hypermultiplet moduli space ${\cal M}_H$.
Just as in six dimensions, tuning a certain number of hyper 
multiplets to special values  leads to a perturbative
gauge symmetry enhancement, but
now also with enlarged rank. 
This effect opens the possibility to perform perturbative Higgs transitions
when going to the Coulomb phase of the newly obtained gauge group.
During this Higgs transition $N_V$ increases, whereas $N_H$ is reduced.
Of course also the reverse Higgs transition is possible.
For example, if we start from the three models with
a completely broken gauge group ($k=0,1,2$)
a perturbative $SU(2)$ can be opened from the original $E_8^1$ by tuning 
$12 k+29$ hyper multiplets to special values. Then, via the 
Higgs transition we
can reach heterotic vacua with $N_V=4$ and $N_H=215-12 k$.

So in conclusion, a large number of four dimensional heterotic staring vacua
can be obtained 
\cite{candelas,AFIQ}, 
via Higgs transitions in the heterotic gauge group $G_1\times
G_2$. For example, consider again the embedding of $(12+k,12-k)$
instantons in $E_8\times E_8$, and perform Higgs transitions in the
first gauge group factor $G_1$ (i.e. assume that $G_2$ is still broken as
far as possible). Then the four-dimensional spectra
$(N_V,N_H)$ of perturbative heterotic string vacua 
can be computed by the  techniques described above
and
are summarized in the 
following two tables \cite{candelas} (in the first horizontal lines,
the unbroken groups $G_1$ are denoted, and the first vertical line shows the
different values $k$): 
\vskip0.2cm
\begin{center}
\begin{tabular}{|c||c|c|c|c|c|c|c|c|} \hline
  & $1$&$SU(2)$ & $SU(3)$&$SU(4)$&$SU(5)$&$SO(10)$&$E_6$&$E_7$ \\ \hline\hline
0 &(3,244)&(4,215)&(5,198)&(6,183)&(7,168)&(8,165)&(9,160)&(10,153)
 \\ \hline
1 &(3,244)&(4,203)&(5,180)&(6,161)&(7,143)&(8,139)&(9,133)&(10,125)
 \\ \hline
2 &(3,244)&(4,191)&(5,162)&(6,139)&(7,118)&(8,113)&(9,106)&(10,97)
 \\ \hline
4 &(7,272)&(8,195)&(9,154)&(10,123)&(11,96)&(12,89)&(13,80)&(14,69)
 \\ \hline
6 &(9,322)&(10,221)&(11,168)&(12,129)&(13,96)&(14,87)&(15,76)&(16,63)
\\ \hline
8 &(10,377)&(11,252)&(12,187)&(13,140)&(14,101)&(15,90)&(16,77)&(17,62)
\\ \hline
12&(11,492)&(12,319)&(13,230)&(14,167)&(15,116)&(16,101)&(17,84)&(18,65)
\\
\hline
\end{tabular}
\end{center}
\vskip0.2cm

\noindent {\bf Table 2:} Perturbative heterotic spectra in 4 dimensions.

\vskip0.2cm
\noindent Of course this list could be easily completed
(making a three-dimensional plot) by considering also the
possible Higgs transitions in the second gauge group factor.


\subsubsection{Non-perturbative Effects}

In this paragraph we will discuss a few non-perturbative effects which 
 play an important role
 at singular points in the heterotic moduli spaces, where
non-perturbative BPS states become massless.
In particular we first want to adress the question how non-perturbative
transitions beween heterotic vacua with different $E_8\times E_8$ instanton
numbers $(n_1,n_2)$ can occur, i.e. how non-perturabtive
effects allow for vertical transitions in table (2)
(Remember that the horizontal transitions are realized by the perturbative
Higgs transitions.)
Then we want to discuss singularities at strong heterotic coupling and the
related issue of strong coupling transitions to non-perturbative
heterotic vacua with frozen dilaton field. Finally we will discuss
non-perturbative $S$-duality symmetries which are expected to be 
present in some
particular heterotic models.

We will start the discussion in six dimensions. Singularities in the heterotic
moduli spaces can either occur at arbitrary, i.e. also at weak, or at
strong heterotic coupling. Considering the first possiblity, it can happen
\cite{witt2} that
at some special locus
in  the hypermultiplet moduli space 
${\cal M}_H$ one or several of the gauge
instantons shrink to zero sizes such that the gauge bundle becomes singular.
This singularity occurs at arbitrarily weak coupling due to the absence
of communication between the hyper and the tensor moduli fields, but
nevertheless it is a non-perturbative effect which has no conformal
field theory description. The corresponding singularity in the efective
field theory is caused by BPS states which become `massless' at the special
locus. For the case of the $SO(32)$ heterotic string the extra BPS states
are given by non-perturbative gauge fields which become massless when an
instanton shrinks to zero size \cite{witt2}.
In this way the rank of the gauge group can be enhanced non-perturbatively
beyond the bound set by conformal field theory.
On the other hand, for the case of the $E_8\times E_8$ heterotic string it was
argued that in general no  vector fields become massless 
at the  loci of zero size instantons,
but a non-critical string becomes tensionless \cite{seiwi}, where
the tension of this string is controlled by the vevs of the hypermultiplets.
The world sheet description of this non-critical string is characterized
by (0,4) world sheet supersymmetry which supports (0,1) supersymmetry
in the six-dimensional target space.
On the world sheet of the tensionless string there lives a massless
antisymmetric tensor field $B_{\mu\nu}^+$; hence at the special locus
of ${\cal M}_H$ there will be new massless tensor multiplets in six
dimensions.
The gravitational anomaly constraint eq.(\ref{gravmatch}).
implies that in order to get one extra massless tensor, i.e.
to shrink one $E_8$ instanton to zero size, 
29 hypermultiplets have to be tuned to special values.
The emergence of the additional tensor multiplets can be also seen from the
Green-Schwarz anomly condition eq.(\ref{gsmatch5}). Namely, 
shrinking one $E_8$ instanton to zero size,
i.e. $n_i\rightarrow n_i-1$, equation (\ref{gsmatch5}) 
demands that $n_5\rightarrow n_5+1$,
i.e. we get one extra massless tensor. 
So this transition leads to a new non-perturbative phase with extra
tensor multiplets, and the new non-perturbative Coulomb branch is parametrized
by vevs of the additional scalars in the tensor multiplets.

Now it also becomes rather clear how the non-perturbative transitions between
different perturbative vacua with different instanton numbers $(12+k,12-k)$
can be realized. First one goes to the special locus, where, say, one
$E_8^1$ instanton becomes small; so $n_1\rightarrow 11+k$. At the second
step one blows up one instanton in $E_8^2$, i.e. $n_2\rightarrow 13-k$.
The net-effect of these two transitions with an intermediate
non-perturbative phase is clearly given by the change $k\rightarrow k-1$.
The (intermediate) Coulomb branch with non-perturbative tensor fields
has no geometrical description in terms of the uncompactified ten-dimensional
heterotic string theory. Instead the transition between different perturbative
vacua has a very nice explanation 
\cite{seiwi} in $M$-theory, compactified on $K3\times
S^1/{\bf Z}_2$. The two $E_8$ gauge factors are not sitting in the bulk of
11-dimensional space time, but are on two different  ``end of the world"
  9-branes \cite{horavawitten}. 
The location of the $n_5$ 5-branes, which fill 6-dimensional space-time,
on $K3\times S^1/{\bf Z_2}$ are labeled by five real parameters, namely
one hypermultiplet (the location on $K3$) and a tensor multiplet
(the coordinate on $S^1/{\bf Z}_2$).
The singularity from  the zero size $E_8$ instanton can be interpretated
as resulting from a 5-brane stuck to the 9-brane. Then in the new Coulomb phase,
the 5-brane leaves the boundary, so the instanton is converted into a 5-brane
which is emitted into the bulk. The 5-brane can now move to the other
end 9-brane and can be converted into an instanton again.
So in this way all values of $(n_1,n_2)$ are connected in $M$-theory.

In four dimensions this effect describes a transition to a non-perturbative
Coulomb branch with additional massless non-perturbative vector multiplets.
The effective interactions of the non-perturbative vector fields are
given by special Chern-Simons couplings, which can be also described
by Poincare dualizing the vector multiplets to $N=2$ four-dimensional
vector-tensor multiplets. Models
with several non-perturbative vector
multiplets  and their  couplings have been discussed in \cite{LSTY,dewitfaux}.
In the next section we will discuss how these transition, which are
non-perturbative on the heterotic side, are realized on the dual type II side
by a change in the topology of the underlying Calabi-Yau space.

Now we like to adress the question
whether it is possible to obtain  heterotic string compactications with no
perturbative tensor multiplets in six-dimensions at all, 
i.e. with no $S$-field
in four dimensions? This question is related to the second type of 
singularity,
which occurs only at strong coupling, i.e. for some special value of the
dilaton field. So let us investigate the nature of the strong coupling
singularity on more detail.
Namely it is very closely related to the gauge kinetic term in six
dimensions, given as
\begin{equation}
{\cal L}_{\rm kin}=-\sum_a(v_ae^{-\phi^6/2}+\tilde v_ae^{\phi^6/2})
{\rm tr}F_a^2,\label{gaugekin6}
\end{equation}
where the second term proportional to $\tilde v_a$ is a one-loop
contribution. Since $\tilde v_a$
can be negative (the tree level term $v_a$ is always positive),
it  follows that at the critical string coupling
\begin{equation}
e^{\phi^6_c}=-{v_a\over\tilde v_a}\label{critdil}
\end{equation}
the gauge coupling of one gauge group factor diverges.
In this case of infinite six-dimensional gauge coupling one deals with
non-trivial gauge dynamics in the infra-red; as it was argued in \cite{seiwi}
the singularity is again related to a non-critical string becoming
tensionless. In fact in six dimensions, the (0,1) supersymmetry algebra
allows for a vector central charge carried by BPS strings which couple
electrically or magnetically to the $B$ field. For a single
$B$-field, the tension obeyes the
BPS bound
\begin{equation}
T\geq |n_e e^{\phi^6/2}+n_me^{-\phi^6/2}|,\label{tension}
\end{equation}
which approaches zero for the 
critical coupling $e^{\phi^6_c}=-{n_m\over n_e}$.

Let us see what is happening at the critical couplings $\phi^6_c$,
starting from the perturbative vacua with $E_8\times E_8$ instanton numbers
$(12+k,12-k)$.
In these models 
the ratios $\tilde v_a/v_a$ are determined as
\begin{equation}
{\tilde v_1\over v_1}=-{\tilde v_2\over v_2}={k\over 2}.\label{v12}
\end{equation}
Assuming $k\geq 0$, the gauge coupling of $G_1$ is always finite,
whereas the other gauge coupling diverges
at
\begin{equation}
e^{-\phi^6_c}={k\over 2}.\label{critdilk}
\end{equation}
Compactifying further to four dimensions on $T^2$ with K\"ahler modulus
$T$, the singularity occurs at the line
\begin{equation}
S={k\over 2}T.\label{stk}
\end{equation}
Let us briefly analyze \cite{seiwi} for which values of $k$ there 
could by a strong coupling
transition to a Higgs branch with no tensor multiplet at all. 
In such a branch there is only the anti-self-dual $B_{\mu\nu}^-$ field
in the supergravity multiplet, and anomaly cancelation requires that the 
anomly 8-form $I_8$ in eq.(\ref{gravan}) not only factorizes
but must be a perfect square.
Using the expressions eq.(\ref{v12}) for $\tilde v_a/v_a$ this is only possible
for $k=1,4$. For other values of $k$ there is no Higgs branch without
dilaton.  The $k=4$ model, i.e. the model with instanton distribution
$(16,8)$, has unbroken (minimal)
gauge group $SO(8)$ in six dimensions; 
this model is known \cite{many} to be $T$-dual to the $SO(32)$
heterotic string with instanton number $n=24$.  

Finally we now discuss the presence of non-perturbative
$S$-duality symmetries in $N=2$ heterotic models. 
First consider the $k=0$ model with symmetric 
$(12,12)$ instanton embedding. 
If we un-Higgs some part of the gauge group, we get $\tilde v_1=\tilde v_2=0$;  
it is
obvious that there is no strong coupling singularity for this model.
So one can continuously extrapolate from weak coupling
to strong coupling without hitting
any phase transition point. 
This suggests that there might be 
a strong-weak coupling $S$-duality
for this model. Indeed going to the $M$-theory description
of the heterotic string one can strongly argue 
that this model has a heterotic
strong-weak coupling $S$-duality symmetry \cite{dmw}
\begin{equation}
\phi^6\leftrightarrow -\phi^6.\label{sdual6}
\end{equation}
The dual heterotic string comes from wrapping the heterotic 5-brane around
$K3$ which is consistently only possible for symmetric instanton embedding.
However, since $v_a\neq 0$, $\tilde v_a=0$ there is no manifest self-duality,
but the heterotic strong-weak coupling duality requires the existence
of non-perturbative gauge bosons which have $v_a=0$ and $\tilde v_a\neq 0$.
So the heterotic strong-weak coupling duality of this model exchanges
perturbative with non-perturbative gauge fields. The enhancement
loci in ${\cal M}_H$ of the perturbative and non-perturbative gauge bosons
are symmetric but not identical. Therefore the strong-weak coupling duality
acts on the hyper multiplet moduli in a non-trivial way, i.e. the
$K3$ gets changed under the duality transformation.
Remember that for the (12,12) model one has to tune 29 hypermultiplets
to open a perturbative $SU(2)$ gauge group. So it takes the same number
of parameters to open an perturbative $SU(2)$ gauge group as 
is needed to shrink
an $E_8$ or $SO(32)$ instanton. This observation again supports the
strong-weak coupling $S$-duality of this model;
the appearance of non-perturbative gauge bosons in case of small $E_8$
instantons
 is naturally expected since the $E_8\times E_8$
heterotic string with symmetric instanton embedding is $T$-dual
to the $SO(32)$ type I string where zero size instantons lead to a
non-perterturbative $SU(2)$ gauge symmetry.

Upon further compactification to
four dimensions on $T^2$ the strong-weak coupling duality eq.(\ref{sdual6})
becomes a non-perturbative
symmetry under the exchange of the $S$ and the $T$ field:
\begin{equation}
S\leftrightarrow T
\end{equation}
The presence of this non-perturbative 
exchange symmetry is in contrast to the $N=4$ supersymmetric vacua
where, as discussed in section (2.1), the exchange of $S$ and $T$
is not a symmetry of the heterotic string but
maps the $N=4$ heterotic string onto the dual type II string.
 In the dual $N=2$ type II models, the non-perturbative
$S$-$T$ exchange
symmetry will find a nice geometrical explanation by considering the
corresponding Calabi-Yau spaces.
The $S$-$T$ exchange symmetry implies that for $T\rightarrow\infty$ there
is a modular symmetry $SL(2,{\bf Z})_S$. Moreover, since in the
perturbative region $S\rightarrow\infty$ there was a perturbative gauge
symmetry enhancement at $T=U$, there is a non-perturbative
$SU(2)$ enhancement at the line $S=U$ for $T\rightarrow\infty$ \cite{CCLMR}.

Finally let us discuss the model $k=2$, i.e. with instanton embedding
$(14,10)$ which also allows for complete Higgsing of the gauge gauge group.
We will see that this model is in fact very closely 
related to the previously studied
case with symmetric (12,12) instanton embedding. For $k=2$, we get
$\tilde v_2/v_2=-1$; so the possible strong
coupling singularity is at $\phi^6_c=0$.
However this singularity is non-generic since it requires in addition to
tune also one hypermultiplet modulus field. Therefore there is no strong
coupling transition possible at this point. The reason for the absence of a 
transition is that the tensionless string which appears at the point
$\phi^6=0$ carries (4,4) world sheet supersymmetry and hence supports
(2,0) supersymmetry in the target space in contrast to the cases considered
before (though the full space-time configuration has still (0,1)
supersymmetry). These  non-critical (4,4) strings 
become tensionless  if one tunes
five real parameters, i.e. one tensor multiplet plus one hypermultiplet.
The singularity can be avoided keeping the hypermultiplet at generic values.
The uniqueness of six-dimensional (0,2) supergravity implies the
absence of a phase transition; so no new Higgs branch can emerge  at
the singular locus.
These observations suggest that like the (12,12)
model also the (14,10)  model possesses 
the strong-weak coupling $S$-duality symmetry eq.(\ref{sdual6}), as
it was noted in \cite{afiq}. This symmetry is possible
since the singularity occurs at the self-dual point of the duality
transformation. However for the (14,10)
model  the $S$-duality does not require the existence of non-perturbative
gauge bosons since $v_1=\tilde v_1$. Hence in this sense, the model is
really self-dual.
Compactifing further to four dimensions it immediately follows that the
(14,10) model also possesses a $S$-$T$ exchange symmetry. In four dimensions
at the singular locus $S=T$ there is no tensionless string but a massless
non-perturbative $SU(2)$ gauge boson together with one massless adjoint
$N=2$ hypermultiplet. (This becomes more clear in the dual type II picture.)
Therefore this gauge theory is finite, because the $\beta$-function
coefficient is proportial to $N_V-N_H$.

We have seen that the symmetric (12,12) model and the (14,10) models
have many common features such as the possiblity to Higgs the gauge
group completely, the absence of a strong coupling singularity and the
presence of the non-perturbative $S$-duality symmetry. This
suggest that these two models might be equivalent and perhaps related
by some sort of $T$-duality symmetry. In fact, using the 
dual type II  or the $F$-theory picture of these two models,
it will become evident that they are indeed completely equivalent
after taking into account all non-perturbative effects. There are
on the same moduli space, or more precisely, the moduli
space of the (14,10) model is a subspace of the moduli space of the (12,12)
model. In other words, upon restricting certain moduli of the (12,12) model
to a particular domain, the (14,10) model is fully covered.
It follows that tuning one hypermultiplet, also the (12,12)
model acquires a singularity at the line $S=T$.

\subsection{The type IIA string on $CY^3$}


\subsubsection{$K3$-fibrations}

The compactification of the ten-dimenional type IIA/B superstring
on a Calabi-Yau three-fold leads to string vacua with $N=2$ space-time
supersymmetry in four dimensions. The massless spectrum is determined
by the cohomology of the three-fold, namely by the two Hodge number
$h^{(1,1)}$ and $h^{(2,1)}$.
As it is known, the type IIA/B 
compactifications are related by mirror symmetry, which means
that  the type IIA superstring on the Calabi-Yau space $X^3$
is equivalently described by compactifying the type IIB on the mirror
manifold $\tilde X^3$ which has exchanged Hodge numbers: $\tilde h^{(1,1)}=
h^{(2,1)}$, $\tilde h^{(2,1)}=h^{(1,1)}$.
We will mainly use the type IIA description in the following; however
the type IIB picture is very useful, for example, for computing
the low-energy effective action or for describing the conifold transition.
The massless $N=2$ spectrum 
of the type IIA superstring compactified on
$X^3$ contains first the $N=2$ supergravity multiplet including the $U(1)$
graviphoton field, which orignates from the ten dimensional R-R gauge
field $A_M$. Second, there are $h^{(1,1)}$ $U(1)$ vector multiplets:
\begin{equation}
N_V=h^{(1,1)}.\label{nv11}
\end{equation}
Their $h^{(1,1)}$ complex scalar moduli in the NS-NS
sector correspond to the deformations of the K\"ahler form $J$ of
$X^3$ plus the internal
$B_{MN}$ fields; the $h^{(1,1)}$ $U(1)$ R-R vectors  originate from
the ten-dimensional 3-form gauge potential $A_{MNP}$ with two indices
in the internal space. So the Abelian gauge symmetry including
the graviphoton is given by
$U(1)^{h^{(1,1)}+1}$. Finally there are $h^{(2,1)}+1$ massless $N=2$
hypermultiplets:
\begin{equation}
N_H=h^{(2,1)}+1.\label{nh21}
\end{equation}
 $h^{(2,1)}$ of them correspond to the complex structure
deformations of $X^3$, where the two additional R-R
scalar degrees of freedom, needed to fil an $N=2$ hypermultiplet,
come from $A_{MNP}$ with all indices in the internal direction. The additional
hypermultiplet contains together with the
NS-NS axion field $a$ the four-dimensional dilaton $\phi^4_{IIA}$
plus two  more R-R scalar fields.

It is clear that for a dual heterotic/type IIA $N=2$ string pair
the number of massless states have to match. 
The fact that the IIA dilaton belongs to an hypermultiplet is in
apparent contrast to the heterotic dual where the dilaton belongs
to a vector multiplet. Therefore the heterotic/type II duality relations will
identify a particular internal type II Calabi-Yau modulus,
$t_S$, with the heterotic
dilaton field $S_H$, and in turn one of the heterotic hypermultiplet moduli
with the type II dilaton.
Moreover the effective actions, i.e. the moduli spaces have to agree, i.e.
${\cal M}_V^H={\cal M}_V^{IIA}$, ${\cal M}_H^H={\cal M}_H^{IIA}$.
This will be the main subject of section 3. However let us already mention
that
 the difference in the structure of the
heterotic/type IIA multiplets 
with respect to the dilaton
field, whose vev sets the
string coupling constant, has very important consequences for the 
effective couplings. Namely due to the $N=2$ non-renormalization theorems
which forbid mixings in the kinetic enenergies of the vector and
hypermultiplets, it follows that ${\cal M}_V^{IIA}$ does not receive
any quantum corrections at all, whereas on the heterotic side
${\cal M}_V^H$ receives perturabtive as well as non-perturbative
corrections. This very powerful observation will enable us to determine
non-perturbative corrections to the classical heterotic vector
moduli space  by computing the purely classical space ${\cal M}_V^{IIA}$
of the dual type IIA model. Obviously, the converse statement is true
for ${\cal M}_H$. However  computing the purely classical heterotic
space ${\cal M}_H$ is very complicated, and not many results exist
in this direction.

Now let us discuss what is the general structure of a Calabi-Yau space $X^3$
such that there exist a dual perturbative heterotic vacuum. We have to require
that
in a particular corner of the Calabi-Yau K\"ahler moduli space, 
which corresponds to the limit $S_H\rightarrow\infty$, we must
see all known classical heterotic effects, like the classical moduli
space
eq.(\ref{mv}), the target space duality symmetries, the classical enhancement
of the Abelian gauge group at special points in the moduli space and,
finally, the possiblity to perform Higgs transitions 
through points of enhanced
gauge symmetries. 
Let us start with the information about $X^3$ we can deduce
from the form of the classical special K\"ahler moduli space eq.(\ref{mv}).
Consider the cohomology of $X^3$ and introduce a basis of $H_4(X^3,{\bf Z})$,
the 4-cycles $D_A$, $A=1,\dots, h^{(1,1)}$. Dual to these 4-cycles
we have a basis of 2-forms, $E_A$, generating $H^2(X^3,{\bf Z})$.
Then we expand the K\"ahler form $J$ and the internal $B$-field as
\begin{equation}
iB+J=\sum_{a=1}^{h^{(1,1)}}t_AE_A,\label{kahlermod}
\end{equation}
where the complex numbers $t_A=iB_A+J_A$ denote the complex K\"ahler moduli.
The geometrical moduli space ${\cal M}_V$ is bounded by the wall of the
K\"ahler cone, $\sigma(K)=\lbrace\sum_AJ_AE_A|J_A>0\rbrace$.
We will take $t_1$ to correspond to the heterotic dilaton field $S$, i.e.
$t_1=t_S$.
Next consider the classical cubic `Yukawa' couplings $C_{ABC}$ 
between the K\"ahler
moduli $t_A$. They are given in terms of the classical intersection
numbers among the 4-cycles:
\begin{equation}
C_{ABC}=\sharp(D_A\cap D_B \cap D_C).\label{intersecno}
\end{equation}
As we will explain more in section 3 the intersection numbers determine
the metric of the special K\"ahler moduli space ${\cal M}_V$, in the
limit where the K\"ahler form of $X^3$ is large, i.e. in the limit
$\alpha '/R^2\rightarrow 0$. 
This is the limit where the world-sheet instantons do not give any
contribution to the moduli space metric.
The essence of this discussion is that we can 
can read off from the classical moduli space eq.(\ref{mv}) those topological
intersection numbers which give the dominant contribution in the limit
$t_S\rightarrow\infty$. The result is
\begin{eqnarray}
&~& C_{111}=0, \nonumber\\
&~& C_{11a}=0, \quad a=2,\dots , h^{(1,1)}\nonumber\\
&~& C_{1ab}=\eta_{ab}, \quad a,b=2,\dots ,h^{(1,1)}\label{k3fibr}
\end{eqnarray}
where $\eta_{ab}$ is a matrix of non-zero determinant and signature
$(+,-,\dots ,-)$. 
Without going into the details of the proof \cite{asplou}
this  specific form of the intersection number implies that the Calabi-Yau space
$X^3$
must be $K3$ fibration, $X^3\rightarrow P^1$, where the $K3$ is varying
over a two-sphere $P^1$.\footnote{$X^3$ can also be a $T^4$
fibration over $P^1$.}
The K\"ahler class, i.e. the size of the base $P^1$, is just given
by the modulus $t_S$. This means that the weak coupling limit on the
heterotic side corresponds on the type IIA side
to the limit where the  $K3$ is fibred over a large base space $P^1$.
This limit looks like a decompactification to six dimensions, i.e.
type IIA on $K3$, but with the difference that only part of the spectrum
survives after the fibration.
Now it also becomes clear that the modular target space duality symmetries
arise in the type IIA compactifications in the limit of large $P^1$, since
the $K3$ possesses modular properties as discussed in section (2.1).

Knowing that $X^3$ is a $K3$ fibration, we now like to analyze more closely
the second cohomology $H^2(X^3)$ or equivalently
$H_4(X^3)$; $h^{(1,1)}$ gets contributions
from the cohomology of the base $P^1$ as well as from the cohomology of the
fibre $K3$: 
\begin{equation}
h^{(1,1)}=1+\rho+\sharp({\rm degenerate}~{\rm fibres}).\label{h11}
\end{equation}
Let us discuss the three terms which appear in this formula.

\noindent (i) First the K\"ahler class $t_S$ will contribute to $h^{(1,1)}$.

\noindent (ii) Next consider a 2-cycle of the fibre $K3$ and ``sweep out"
a 4-cycle in $X^3$ by transporting it around the base $P^1$.
This can be done in a consistent way if the $K3$ 2-cycle is monodromy 
invariant. The number of monodromoy invariant 2-cycles in $K3$
is given by the socalled Picard number $\rho$, where $\rho$ depends on 
how the
$K3$ is fibred over the $P^1$. For Calabi-Yau spaces whose heterotic
dual comes from a $K3\times T^2$ compactification, it follows that
$\rho\geq 2$, where
the three universal K\"ahler moduli always correspond to the heterotic
fields $S$, $T$ and $U$. In this case $X^3$ is also an elliptic fibration
over a  complex two-dimensional space $B^2$ (see the $F$-theory section).

\noindent (iii) There can be a degenerate fibre which is a reducible divisor
in $X^3$. It turns out that these classes cannot be understood in 
perturbative
heterotic string vacua. 

Now consider very briefly the complex structure deformations of $X^3$.
First,  $h^{(2,1)}$ receives contribitions from varying the
complex structure of the $K3$ fibre; this number is given by $20-\rho$.
So we see that with increasing number of monodromy invariant two cycles
the $K3$ becomes more and more rigid, i.e. it provides less complex structure
deformations. Second the deformations of the fibration contribute to
$h^{(2,1)}$. So in total one gets
\begin{equation}
h^{(2,1)}=20-\rho+\sharp({\rm fibration}~{\rm deformations}).\label{h21}
\end{equation}

In the limit $t_S\rightarrow\infty$  one  recoveres
the perturbative gauge symmetry enhancement of the heterotic string
at special points in the moduli space \cite{aspinwall}.
Since we have to take the large $P^1$ limit, we are dealing almost with
the $N=4$ situation of type IIA compactification on $K3$. 
The charged gauge bosons are again given by non-perturbative
D 2-branes which can wrap around the 2-cycles of $K3$.
They become massless if the $K3$-fibre degenerates by shrinking the
sizes the 2-cycles.
However the  difference compared to the $N=4$
situation  comes from  possible global
obstructions due to the $K3$ fibration. Namely
  the wrapping of D 2-branes
only around
monodromy invariant 2-cycles can lead to physical BPS states.
This means that not the whole intersection lattice
$\Lambda$ of $K3$ can lead to
a non-Abelian gauge group enhancement, but only the monodromy invariant part
of $\Lambda$. This is the socalled Picard lattice $\Lambda_\rho$,
whose rank is given by the Picard number $\rho$. Only $\Lambda_\rho$ 
corresponds to the visible gauge group in case the $K3$-fibre
degenerates. The Picard
lattice has signature $(+,-,\dots ,-)$, and
$\eta_{ab}$ is simply the natural inner
product of $\Lambda_\rho$.

After having discussed the general structure of the Calabi-Yau spaces which
are dual to perturbative $N=2$ heterotic string vacua, let us now consider
explicit examples of $K3$ fibrations and try to identify them with the
heterotic models which we have discussed in the previous section.
In the following, we
will mainly concentrate on those Calabi-Yau spaces which can be
represented as hypersurfaces in weighted projected spaces.
Consider first  the four-dimensional heterotic vacua with
$k=2,4,6,8,12$ and completely Higgsed gauge group (first column in table 2.)
The numbers $(N_V,N_H)$ of these models are precisely matched by the
Hodge numbers of the $K3$ fibrations given as hypersurfaces of degree
$6k+12$ in $WP_4^{(1,1,k,2k+4,3k+6)}(6k+12)$ \cite{candelas}.
The corresponding $K3$ fibres can also be represented as hypersurfaces
of degree $3k+6$ in $WP_3^{(1,{k\over 2},k+2,{3\over 2}k+3)}(3k+6)$.
The Picard numbers of these $K3$'s are given by $\rho=2,6,8,9,10$
for $k=2,4,6,8,12$ respectively.
Un-Higgsing $SU(r)$ ($r=2,3,4$) gauge group factors (see columns 2 - 4
in table 2), the dual Calabi-Yau space can be again given by hypersurfaces
in weighted projected spaces. 

In summary,
the  Calabi-Yau models  $X^3$ which can
be represented as hypersurfaces in weighted projective spaces and
which are the type IIA
duals to the heterotic compactifications with $k=2,4,6,8,12$ and $r=1,2,3,4$
are listed in the following table.\footnote{The
exist many more $K3$-fibred Calabi-Yau spaces (see  \cite{klm,manymore}).
Some of them could be shown to possess a dual
heterotic vacuum ($B,C$ chains) which is however not
a $K3\times T^2$ compactification with $(12+k,12-k$)
instanton embedding.} The corresponding Hodge numbers
can be read off from the heterotic spectra $(N_V,N_H)$ using eqs.(\ref{nv11})
and (\ref{nh21}).

\vskip0.2cm
\begin{center}
\begin{tabular}{|c||c|} \hline
  $r$     &  $X^3$  \\ \hline\hline
1  &   $WP_4^{(1,1,k,2k+4,3k+6)}(6k+12)$   \\ \hline
2 &    $WP_4^{(1,1,k,k+4,2k+6)}(4k+12)$ \\ \hline
3 & $WP_4^{(1,1,k,k+4,k+6)}(3k+12)$  \\ \hline
4 & $WP_5^{(1,1,k,k+4,k+6,k+8)}(2k+8,2k+12)$  \\ \hline
\end{tabular}
\end{center}
\vskip0.1cm

\noindent {\bf Table 3:} Dual Calabi-Yau spaces as hypersurfaces in weighted
projected spaces.

\vskip0.2cm
\noindent
So far we are  missing the type IIA duals of many of the heterotic
spectra in table 2, namely all models with $k=0,1$ and the models
with larger un-Higgsed gauge group than $SU(4)$. These models cannot be
simply represented by an hypersurface in a weighted projective space; however 
they can be constructed in terms of reflexive polyhedra using the 
techniques of toric geometry \cite{candelas}. In addition, all Calabi-Yau's
dual to the heterotic models with completely
Higgsed gauge group can be written as an elliptic
fibration in the Weierstra{\ss} form 
(see next subsection).

To be specific let us discuss the Calabi-Yau space $WP_4^{(1,1,2,8,12)}(24)$
in more detail \cite{hosono,KV,klm}. 
The corresponding Hodge numbers are $h^{(1,1)}=3$ (i.e. $\rho=2$) and 
$h^{(2,1)}=243$. 
The three K\"ahler moduli are 
$t_1$, $t_2$ and $t_3$. This  IIA model 
is dual to the heterotic string with instanton
embedding $(n_1,n_2)=(14,10)$ and completely Higgsed gauge group. The
three heterotic 
 vector fields were denoted by $S$, $T$ and $U$. 
 Therefore this model is called 
the $S-T-U$ model.
The identification between the Calabi-Yau K\"ahler parameters and the
heterotic moduli is given as (see the next chapter about the prepotential):
\begin{equation}
t_1=S-T,\quad t_2=U, \quad t_3=T-U.\label{stmapk2}
\end{equation}
The positivity of the K\"ahler cone implies that the heterotic variables are
ordered as $S>T>U$.

The defining polynomial of the Calabi-Yau space is
\begin{equation}
p={1\over 24}(z_1^{24}+z_2^{24}+2z_3^{12}+8
z_4^3+12z_5^2)
- \psi_0z_1z_2z_3z_4z_5
 - {1\over 6}\psi_1(z_1z_2z_3)^6 -{1\over 12}\psi_2
(z_1z_2)^{12}.\label{polynomial}
\end{equation}
In this equation we have included three deformation parameters $\psi_0$,
$\psi_1$ and $\psi_2$ which denote three of the complex structure deformations
of this Calabi-Yau; 
furthermore  there are 239 additional allowed deformations 
which could be added to $p$; however one complex structure deformation
of $X^3$ cannot be represented by a deformation of this defining polynomial.
In the type IIB compactification on the mirror space $\tilde X^3$ with
$h^{(1,1)}=243$ and $h^{(2,1)}=3$ the defining polynomial $p$
contains all possible allowed deformations. The relation between the 
the K\"ahler parameters $t_A$ of the type IIA model on $X^3$ and
the complex structure parameters $\psi_A$ of the type IIB model on 
$\tilde X^3$ can by found by using the mirror symmetry between these two
three-folds \cite{hosono}.

At special points in the moduli space the Calabi-Yau degenerates and acquires
a conifold singularity. The degenaration takes place on the locus where
the Calabi-Yau discriminant $\Delta$ is vanishing:
\begin{equation}
\Delta=(y-1) \times {(1-z)^2-yz^2\over z^2}  \times 
{((1-x)^2-z)^2-yz^2\over z^2} =
\Delta_y \times \Delta_z \times \Delta_x;\label{disclocus}
\end{equation}
here $x=-\psi_0^6/\psi_1$, $y=1/\psi_2^2$ and $z=\psi_2/\psi_1^2$
are functions of the three vector fields $S,T,U$.
For $y\rightarrow 0$, $\Delta$ degenerates into quadratic factors that will
provide the loci where precisely the perturbative heterotic gauge symmetry
enhancement takes place.
Therefore it is natural to make the identication
$y=e^{-2\pi S_{\rm inv}}$, where $S_{\rm inv}$ is a redefined dilaton field
which is invariant under the target space duality transformations \cite{DKLL}
(see also next chapter). 
Moreover it turns out that in the weak couling limit $y=0$ the mirror map
can be explicitly solved, and $x$ and $z$ can be written in terms of the
elliptic $j$-function as 
\begin{eqnarray}
x &=& {1\over 864}\Biggl({j(T)j(U)\over
j(T)+j(U)-1728} +{\sqrt{j(T)j(U)(j(T)-1728)(j(U)-1728)}\over
j(T)+j(U)-1728}\Biggr),\nonumber\\
z&=&864^2{x^2\over j(T)j(U)}.\label{xzfunc}
\end{eqnarray}
So in this limit one recovers
the perturbative duality symmetry
$SL(2,{\bf Z})_T\times SL(2,{\bf Z})_U\times {\bf Z}_2^{T\leftrightarrow U}$
from the modular properties of the elliptic $j$-function,
as we have expected since $X^3$ is a $K3$-fibration.
Using the expressions eq.(\ref{xzfunc}), we can easily recover
the vanishing of the discriminant locus $\Delta$
at the lines of classical gauge symmetry enhancement.
First at the line $T=U$, i.e. $j(T)=j(U)$, $z=1$ and $x={1\over 864}j(T)$
and hence $\Delta_z=0$ and 
${\sqrt\Delta_x}={4j(T)(j(T)-1728)\over 1728^2}$.
The points $x=0$ ($T=e^{i\pi /6}$), $x=2$ ($T=1$),
where $\Delta_x=0$, correspond to the points
of further enhanced gauge symmetries $SU(3)$ or $SU(2)^2$ respectively.

Let us now briefly discuss what is happening 
if we go to finite coupling, i.e. taking
into account quantum corrections on the heterotic side.
For finite $y$ the quadratic degenerations of $\Delta_x$ and
$\Delta_z$ are lifted and
the classical loci of enhanced gauge symmeries split into two separated
singular lines. Moreover, the classical target space duality symmetries
disappear for finite couplings. In particular the Weyl reflections,
like the $T\leftrightarrow U$ exchange, are not any more quantum
symmetries.
These effects are precisly what is happening in $N=2$ supersymmetric
field theories, as discovered by Seiberg and Witten \cite{sw}.
The  singular lines at $\Delta_x=0$ or $\Delta_z=0$,
where the Calabi-Yau has a conifold
singularity for finite $y$, correspond to the loci of massless monopoles and
dyons. 
So there is no gauge symmetry enhancement at $T=U$ in the quantum case,
but at $T-U=\pm \Lambda$ one gets massless monopoles or dyons.
In fact one can verify that the perturbative monodromies from the duality
transformations split into monopole times dyon like monodromies \cite{clm1},
and that
(in a somewhat simpler model with $h^{(1,1)}=2$ and only
classical $SU(2)$ gauge symmetry enhancement) 
that the monodromies around the conifold
points precisely agree with the monodromies around the points of massless
monopoles and massless dyons after performing suitable field theory limit
$\alpha'\rightarrow 0$ \cite{kklmv}. 
Moreover, considering only the local geometry around
the singularities, the full Seiberg-Witten curves can be reconstructed.
In a type IIB picture the conifold points are described by those loci
in ${\cal M}_V$ where certain 3-cycles shrink to zero sizes.
The massless monopoles and dyons of Seiberg and Witten correspond to BPS
self-dual 3-branes which are wrapped around the singular 3-cycles.
In the mirror type IIA models, the Seiberg-Witten effect is then 
translated into the wrapping of 2-branes around isolated  2-cycles which 
shrink
to zero with respect to the quantum ($\alpha'$) corrected metric.
In this way, string theory provides a very beautiful geometrical
description of non-perturabtive effects in field theory, an approach
which is called `Geometrical Engineering of Field Theories" \cite{engin}.

The $S$-$T$-$U$ model possesses a further singularity at $y=1$ where
$\Delta_y=0$. This locus does
not intersect with the Seiberg-Witten locus, and
the corresponding singularity cannot be found in field theory.
This Calabi-Yau singularity corresponds to the
strong coupling singularity of the heterotic (14,10) model at the
line $S=T$, which we have discussed before. The geometric singularity
is a genus one curve of $A_1$ singularity \cite{kmp}, predicting
that $SU(2)$ gauge bosons plus an adjoint hypermultiplet become
massless at the singularity. 
To reproduce this singularity on the Calabi-Yau space,
one has to go to a locus of codimension one in ${\cal M}_V$
($S=T$), but in addition also to  a locus of codimension one in
${\cal M}_H$, namely to the codimension one space of those complex
structures which correspond to the polynomial deformations of $p$.
This 
strong coupling effect  has no field theory explanation and is therefore
an entirely new stringy effect. Analyzing further the symmetries of this
Calabi-Yau compactification it turns out \cite{klm} that,
instead of the classical $T$-dualities, the model is symmetric
under the quantum symmetry $t_1\rightarrow -t_1$, $t_3\rightarrow t_1+t_3$.
This symmetry is just the non-perturbative exchange
symmetry $S\leftrightarrow T$. 
It means that this Calabi-Yau has two equivalent $K3$-fibrations 
\cite{asgr},
where the base $P^1_S$ gets exchanged with the $P^1_T$ that is inside
the $K3$, regarded as an elliptic fibration over $P^1_T$.

After having discussed this specific Calabi-Yau model in some detail
we now want to investigate how the perturbative heterotic  Higgs transitions 
are realized in the dual type II Calabi-Yau models. 
As explained, such transitions
go through enhanced non-Abelian gauge symmetries with charged 
matter fields.
When the charged spectrum is such that the theory contains a Higgs branch
with  broken gauge symmetry, the transition describes a topology change
to the moduli space of another Calabi-Yau manifold. These transitions
are in general called extremal transitions which generalize the
notion of the conifold transitions.
In the type IIA compactifications $h^{(1,1)}$ gets reduced whereas $h^{(2,1)}$
increases. (We called this the reverse Higgs transition in the heterotic
section.) These transition can occur at weak coupling using perturbative
non-Abelian gauge groups or also at strong coupling when a non-perturbative
non-Abelian gauge groups appears as we discussed in the last paragraph.
In the type IIA picture such transitions occur at the boundary of the
(quantum) K\"ahler cone where a 2-cycle shrinks to zero size.
In the mirror type IIB picture, if a 3-cycle shrinks to a curve (rather
than two a point which leads to a massless hypermultiplet) 
one obtains enhanced gauge symmetries. More precisely,
if the local geometry is an ALE space 
with $A_N$ singularity over a curve of genus
$g$ one obtains an enhanced gauge $SU(N+1)$ symmetry with
$g$ hypermultiplets in the adjoint representation \cite{sadva,kmp,KM}.
As an example \cite{BKKM}, consider the type IIA compactification 
on the Calabi-Yau
$WP_4^{(1,1,2,6,10)}(20)$ with $h^{(1,1)}=4$ and $h^{(2,1)}=190$.
Its heterotic dual is the $K3\times T^2$ compactification with $(14,10)$
instanton distribution and an additional $SU(2)$ from the first
$E_8$ which is generically broken to $U(1)$. By the mechanism described before
this Calabi-Yau has a Higgs transition at weak coupling to the
$S-T-U$ Calabi-Yau $WP_4^{(1,1,2,8,12)}(24)$ with Hodge numbers
$h^{(1,1)}=3$ and $h^{(2,1)}=243$.


\subsubsection{Six-dimensions -- $F$-theory on elliptic $CY^3$}


Now we are interested in how the non-perturbative transitions which
connect heterotic vacua with different instanton number $k$ are
realized on the type II side. In addition we also want to discuss
the dual description of the strong coupling transition to heterotic models
with no dilaton multiplet. As discussed  at the end of section
(2.2) both effects are related to the appearance of tensionless strings
and occur already in the heterotic string on $K3$ in six dimensions
with (0,1) supersymmetry.
Therefore it would be very useful to get a six-dimensional  formulation
of the dual type II strings. One way to achieve this would be perform
a limit in the Calabi-Yau moduli space where the K\"ahler modulus
$t_T$ which corresponds to the heterotic $T$-field gets large.
However this limit is not very easy to analyze. Instead it turns out
to be much more instructive to consider  $F$-theory compactifications
of the type IIB string to six dimensions.

Let us very briefly recall the idea behind $F$-theory.
In \cite{V} Vafa conjectured
 that the II B superstring theory should
be regarded as the toroidal compactification of twelve--dimensional
F--theory. 
If one is slightly less ambitious and does not want to enter
the enormeous difficulties in formulating a consistent 12-dimensional
theory, $F$-theory can be regarded as to provide
new non-perturabtive type IIB vacua
on D--manifolds  
in which the complexified coupling varies over the
internal space. These compactifications then have a beautiful
geometric interpretation as compactifications of $F$-theory on
elliptically fibred manifolds, where the fibre encodes the behaviour of
the coupling, the base is the D--manifold, and the points where
the fibre degenerates specifies the positions of the D 7-branes 
in it. Compactifications of F--theory on elliptic Calabi--Yau
two-folds (the K3), three-folds and four-folds can be argued to be
dual to certain heterotic string theories in 8,6 and 4 dimensions and
have provided new insights into the relation between geometric 
singularities and perturbative as well as non--perturbative 
gauge symmetry enhancement and into the structure of moduli spaces.

Let us consider in more detail $F$-theory on a Calabi-Yau
three-fold $X^3$ which leads to (0,1) supergravity in six dimensions.
This is supposed to be dual to the heterotic string on $K3$.
The Calabi-Yau space $X^3$ is assumed to be an elliptic fibration,
$X^3\rightarrow B^2$, over a two-dimensional base space $B^2$.
The precise fibration data on the $F$-theory side will map to the
gauge bundle data on the heterotic side, in particular to the 
$E_8\times E_8$ instanton numbers.
Upon further compactification to four dimensions on $T^2$ , i.e.
considering $F$-theory on $X^3\times T^2$, we obtain $N=2$
supersymmetric models which are dual 
to the heterotic string on $K3\times T^2$. Using the adiabatic argument one
is then lead to the conclusion that $F$-theory on $X^3\times T^2$ is
equivalent to the type IIA string on the very same Calabi-Yau space
$X^3$.

Let us  discuss \cite{V,MV} how the Hodge numbers 
of the elliptic $X^3$ determine spectrum
of the $F$-theory compactifications. In six dimensions
the number of tensor multiplets $N_T$ is given by the number of
K\"ahler deformations of the (complex) 
two-dimensional type IIB base $B^2$ except
for the overall volume of $B^2$:
\begin{equation}
N_T=h^{(1,1)}(B^2)-1.\label{notensor}
\end{equation}
These tensor fields become Abelian $N=2$ vector fields upon further
$T^2$ compactification to four dimensions. 
Since the four-dimensional $F$-theory
is equivalent to the type IIA string on $X^3$ it follows that the
number of four-dimensional Abelian vector fields in the Coulomb phase
is given by $N_T+r(V)+2=h^{(1,1)}(X^3)$, where $r(V)$ is the rank of the
six-dimensional gauge group, and the additonal two vector fields
arise from the toroidal compactification.
This then leads to the following equation for $r(V)$:
\begin{equation}
r(V)=h^{(1,1)}(X^3)-h^{(1,1)}(B^2)-1.\label{novector}
\end{equation}
Finally, the number of hypermultiplets $H$, 
which are neutral under the Abelian
gauge group, is given in four as well as in six dimensions
by the number of complex deformations of $X^3$ 
plus the freedom in varying the the size of the base $B^2$:
\begin{equation}
N_H=h^{(2,1)}(X^3)+1.\label{nohyper}
\end{equation}

Now we are interested in those $F$-theory compactifications which are
dual to the perturabtive heterotic string vacua on $K3$ with
$n_5=0$ and $(n_1,n_2)=(12+k,12-k)$ $E_8\times E_8$ instanton embedding.
So we expect a one parameters family of elliptic  Calabi-Yau spaces $X^3_k$,
to provide the dual $F$-theory models. At the same time the $X_k$ must
be also $K3$ fibrations over $P^1_b$,
$X^3_k\rightarrow P^1_b$, where the $K3$ fibre itself should
an elliptic fibration over $P^1_{f}$, $K3\rightarrow P^1_f$.
 Putting these observations together
it is suggested that at least locally the (complex) two-dimensional
base space $B^2$ is a product of $P^1_b\times P^1_{f}$. This can be 
formulated \cite{MV} in a more precise way with the result that
$X^3_k$ must be an elliptic fibration over the base $B^2=F_k$, which is
the $k$-th Hirzebruch surface. These surfaces are all $P^1_{f}$ fibrations
over $P^1_b$, and they are distinguished by how the 
$P^1$'s are
twisted. For example, $F_0$ is just the direct product $P^1_b\times P^1_{f}$.
For all $F_k$,
 $h^{(1,1)}(F_k)$ is
universally given by $h^{(1,1)}(F_k)=2$. Therefore one immediately gets
that $N_T=1$, which corresponds to the universal 
heterotic dilaton tensor multiplet
in six dimensions. The Hodge numbers of $X^3_k$ can be read off from the
first column of table 2.

In case of $F$-theory on $K3$, the 8-dimensional heterotic dilaton is just
related to the K\"ahler class (size) $K_f$ of $P^1_{f}$ \cite{V}:
$e^{\phi^8_H}=K_f$. Therefore in six dimensions one gets
\begin{equation}
e^{\phi^6_H}={e^{\phi^8_H}\over K_b}={K_f\over K_b},\label{kfkb}
\end{equation}
where $K_b$ denotes the K\"ahler class of $P^1_b$.
The possible values for the heterotic coupling constant are restricted
by the
boundaries of the K\"ahler cone of the Hirzebruch surface $F_k$.
They lead to the  inequalities $K_f\geq 0$ and
\begin{equation}
e^{-\phi^6_H}={K_b\over K_f}\geq {k\over 2}.\label{boundkfkb}
\end{equation}
This precisely coincides with those values for the string couplings 
(see eq.(\ref{critdilk})) where one gets the strong coupling singularities, and 
tensionless strings appear. In the $F$-theory, i.e IIB context, the
tensionless strings arise by wrapping the D 3-brane around a 2-cycle
which collapses at the boundary of the K\"aher cone. We will come back
to this issue when discussion the strong transitions in the 
$F$-theory picture.
For $k=0$ the compactication is completely invariant under the exchange
of $K_b$ and $K_f$, since we are dealing with a direct product $F_0=P^1_b
\times P^f$, i.e
there is a double $K3$ fibration structure. This symmetry
obviously implies the strong-weak coupling $S$-duality
$\phi_H^6\leftrightarrow -\phi_H^6$.

So far have only determined the rank of the six-dimensional gauge group by
eq.(\ref{novector}). In general the gauge group will contain also 
a non-Abelian part, either generically without tuning moduli (for $k>2$)
or at specific loci (for $k=0,1,2$). The corresponding gauge symmetries
can be determined by analyzing the singularities of the $K3$ fibre
of $X^3_k$.
For that purpose it is very useful to describe the elliptically
fibred Calabi-Yau spaces $X^3_k$ in the Weierstrass form \cite{MV}:
\begin{equation}
X^3_k:\quad 
y^2=x^3+\sum_{n=-4}^4f_{8-nk}(z_1)z_2^{4-n}x
 +\sum_{n=-6}^6g_{12-nk}(z_1)
z_2^{6-n}.\label{weier}
\end{equation}
Here $f_{8-nk}(z_1)$, $g_{12-nk}(z_1)$ are polynomials of degree
$8-nk$, $12-nk$ respectively, where the polynomials with negative degrees
are identically set to zero. From this equation we see that the Calabi-Yau
threefolds
$X^3_k$ are indeed $K3$ fibrations over $P^1_b$ with coordinate $z_1$; the 
$K3$ fibres themselves are elliptic fibrations over the $P^1_{f}$ with
coordinate $z_2$.
The Hodge numbers $h^{(2,1)}(X^3_k)$, 
which count the number of complex structure
deformations of $X^3_k$, are  given by the the number of parameters
of the curve (\ref{weier}) minus the number of possible reparametrizations,
which are given by 7 for $k=0,2$ and by $k+6$ for $k>2$.
The non-Abelian gauge symmetries
are determined 
by the 
singularities of the curve (\ref{weier}) and were analyzed in detail in
\cite{BIKMSV}.
For example, for $k=0,1,2$ it is easy to see that the elliptic curve
(\ref{weier}) is generically non-singular. Only tuning some parameters
of the polynomials $f$ and $g$ to special values, the curve will become
singular.
These $F$-theory
singularities correspond to the perturbative gauge symmetry enhancement
in the dual heterotic models.
On the other hand, for the cases $k>2$ the curve (\ref{weier}) always
contains generic singularities, since on the heterotic side the gauge
group cannot be completely Higgsed. Note that the appearance of massless
matter, i.e massless hypermultiplets also can be analyzed from extra
singularities.

Let us now discuss phase transitions in the Calabi-Yau topology which correspond
to non-perturbative transitions on the dual heterotic side.
Recall that these transitions  occur at singular loci
where the number
of massless tensor multiplets gets changed.
Specifically, if we shrink an $E_8$ instanton to zero size we trade 29
hypermultiplets to get one additional tensor field. 
On the other hand, during the strong transition 
($k=1,4$) we were loosing the
perturbative tensor field, ending up in a vacuum with no tensor field at all. 
Since the number of tensor fields in $F$-theory is entirely related
to the number of K\"ahler parameters of the base $B^2$ (see
eq.(\ref{notensor}))
we learn that these transitions have to due with a change in the
topology of the base $B^2$.
First, to increase the number of K\"ahler parameters of $B^2$ one 
blows up one or several points in $B^2$. For example, blowing up one point
in $F_k$ one obtains one additional tensor field whose scalar vev
controls the K\"ahler class of the blown up.

On the other hand, to loose tensor multiplets one considers extremal
transitions, where the K\"ahler parameters we wish to tune in order
to approach the
transition point are the K\"ahler parameters of the base $B^2$.
In general, extremal transition between  Calabi-Yau spaces take place
going to the boundary of the K\"ahler cone. 
Then
the Calabi-Yau space degenerates in one of the
following three ways \cite{witt3}:
(i) A two cycle collapses to a point.
(ii) A complex divisor (4-cycle) collapses either to  a curve or to a point.
Case (i) corresponds to a topology change via a flop transition
between two different but birationally equivalent
Calabi-Yau spaces with same $h^{(1,1)}$, where 
by moving through the wall a new geometrical
K\"ahler cone is reached; the union of all geometrical K\"ahler cones
is called the extended K\"ahler cone. During the flop transition
one K\"ahler modulus, say $t_2$ changes its sign and the size of the new
2-cycle is given by $-t_2$. This has the effect that the intersection
numbers change by the new term $C_{222}=-{1\over 6}$ \cite{aft}.
(Several concrete examples of flop transitions in particular Calabi-Yau
spaces where investigated in \cite{LSTY}.)\footnote{However
in four-dimensional type IIA compactifications on a Calabi-Yau
space, the flop transition is in fact not a sharp transition since one
can turn on the axionic components of the complex moduli fields. 
Moreover in four-dimensions there are in gereral non-geometrical 
phases outside
the extended K\"ahler cone due to the effect of world sheet
instantons.} Case (ii) on the other hands corresponds to the strong transition
with reduced $h^{(1,1)}$. In case of elliptic Calabi-Yau manifolds 
over $F_k$ the strong transition point occurs where a rational curve ($P^1$)
shrinks to zero size (blow down), and that curve has self-intersection
number $-k$.

We  now can understand how the transition among 
two elliptic Calabi-Yau spaces $X^3_k$ and $X^3_{k\pm 1}$ takes place.
Namely we can go from the base $F_k$ to the new base $F_{k\pm 1}$ by blowing up
one point in $F_k$ and then blowing down one curve of self-intersection
number -1; this corresponds to the shrinking of one $E_8$ instanton and
creating a new instanton in the other $E_8$.

The strong coupling transition to a phase with no dilaton corresponds to
the blowing down of a curve in $F_k$. Specifically for $k=1$ we blow down
a curve of self-intersection number -1, where it turns out that before
the blowing down one has to do a flop transition. This describes the
transition $F_1\rightarrow P^2$. Since $h^{(1,1)}(P^2)=1$, eq.(\ref{notensor})
implies $N_T=0$, in agreement with the strong coupling transition.
Similarly, for $k=4$ the topology change describes the transition 
$F_4\rightarrow P^2$.

\vskip0.2cm

\section{The Effective $N=2$ Action}

So far we have mainly compared the massless spectra and some of the massive
BPS states in dual heterotic/type II string pairs. Now we will
discuss the four-dimensional effective $N=2$ supergravity action of
the
heterotic string on $K3\times T^2$ and of the type IIA superstring on
a Calabi-Yau space $X^3$.
In general a lot of important informations 
about the field dependence of the effective couplings
is gained by studying the
massive BPS spectra of the theory. In particular, it turns out that many
holomorphic Wilsonian 
couplings are BPS dominated, i.e. that they can be computed by integrating
over the massive BPS states \cite{DKL,FKLZ,HM}. At the singular lines 
in the moduli space where some
of the BPS states become massless
(for example massless
gauge bosons), the effective Wilsonian description
breaks down. This breakdown is signalled by singularities in the
field dependend effective couplings at the loci of massless BPS states.
For example, the perturbative Abelian gauge couplings exhibit
logarithmic singularities at the loci of enhanced non-Abelian gauge symmetries,
where the coefficients of the logarithmic terms are just given by the
$\beta$-function coefficients of the enhanced gauge group. This is nothing
else that the known threshold effect in the field theory running
of the gauge coupling constants.
 
Let us briefly recall the structure of the effective couplings with respect to
their dilaton  dependence. 
The form of the field dependent couplings is strongly
constrained by the holomorphy of the Wilsonian part of the effective action.
First,  since the dilaton multiplet $S_H$ is one of the
heterotic vector fields, the holomorphic vector couplings on the
heterotic side do depend on $S_H$;
they
receive classical (of order $S_H^0$), one loop 
(of order $S_H^1$) plus also 
non-perturbative (or order $e^{-S_H}$) contributions.

On the other hand, the dilaton belongs to an hypermultiplet in the 
four-dimensional type II vacua. This implies that the type II vector couplings
(gauge couplings plus moduli space metric)
do not depend on the type II dilaton and are purely classical in string
perturbation theory. Higher derivative 
type II vector couplings will
get perturbative genus $g$ corrections, but no non-perturbative contributions
at all. In type IIA vacua, the vector couplings will receive non-perturbative
world-sheet corrections (of order $e^{-R^2/\alpha '}$), since the
K\"ahler class of the Calabi-Yau space $X^3$ belongs to the vector multiplets.
However in the mirror type IIB models on $\tilde X^3$, the K\"ahler class
is an hypermultiplet, and hence all vector couplings  follow entirely
from classical type IIB geometry. 

Switching to the effective couplings among the $N=2$ hypermultiplets, the
heterotic couplings are classical whereas the type II effective
action will receive perturbative as well as non-perturbative
corrections. However we will not discuss the hypermultiplet effective
action in the following.

The purpose of the comparison of the $N=2$ heterotic/type II effective actions
is two-fold.
First one can obtain
very powerful and non-trivial checks of the $N=2$ string-string
duality for candidate dual string pairs. Considering
the couplings of the $N=2$ vector multiplets, these checks are possible
in the region of small heterotic coupling, $S_H\rightarrow\infty$, where
one can compute the effective heterotic string 
action from string perturbation theory.
Indeed, the string-string duality has been sucessfully tested
by the comparison of lower-order gauge
and gravitational couplings \cite{C,CCLM,CCL}
of the perturbative heterotic string 
with the corresponding couplings of the
dual type-II string.
That is, it was shown that the perturbative heterotic prepotential
${\cal F}$ and the function $F^1$ (which specifies the 
non-minimal gravitational interactions 
involving the square of the Riemann tensor) 
agree with the corresponding type-II functions
in the limit where one specific K\"ahler-class modulus of the
underlying Calabi-Yau space becomes large.  
A set of interesting
relations between certain topological Calabi--Yau data 
(intersection numbers,
rational and elliptic instanton numbers) and various modular forms,
which appear at the perturbative heterotic string,  has
emerged when performing these tests \cite{CCLM,CCL}.

Second the string-string duality can be used to derive the form
of the non-perturbative interactions on the heterotic side from geometrical
considerations on the dual type II side. 
Via the $N=2$ string-string duality,
the topological Calabi-Yau data, in particular the world-sheet
instanton numbers which determine
the {\sl perturbative} type II couplings,
translate into the {\sl non-perturbative} space-time instanton numbers
on the heterotic side. Since the world sheet instanton effects in the type IIA
compactifications can be computed entirely from classical geometry
in the corresponding type IIB vacua by utilizing the the mirror map,
one is calling this map between world-sheet and space-time instantons
the second quantized mirror symmetry \cite{FHSV}.

\subsection{$N=2$ Special Geometry}

The vector couplings of $N=2$ supersymmetric Yang-Mills theory
are encoded in a holomorphic function $F(X)$, where the $X$ 
denote the complex scalar fields of the  
vector supermultiplets. With local supersymmetry this function 
depends on one extra field, in order to incorporate the 
graviphoton. The theory can then be encoded in terms of a 
holomorphic function $F(X)$ which is homogeneous of second 
degree and depends on complex fields $X^I$ with $I=0,1,\ldots 
N_V$. 
An intrinsic definition of special K\"ahler manifold can be given
\cite{c} in terms of a flat $(2n+2)$-dimensional symplectic
bundle over the $(2n)$-dimensional K\"ahler-Hodge manifold, with 
the covariantly holomorphic sections
\begin{equation}
V=\pmatrix{X^I\cr F_I},\quad I=0,\dots ,N_V.\label{section} 
\end{equation}
Usually the $F_I$ can be expressed in terms of a holomorphic 
prepotential $F(X)$, homogenous of degree
two, via $F_I=\partial F(X)/\partial X^I$.
The $N_V$ physical scalar fields
of this system parametrize 
an $N_V$-dimensional
complex hypersurface, defined by the condition that the section satisfies a 
constraint 
$
\langle \bar V,V\rangle \equiv \bar V^{\rm T}\Omega V 
= -i$, 
with $\Omega$ the antisymmetric matrix
$
\Omega = \pmatrix{ 0& {\bf 1} \cr -{\bf 1} &0 \cr }$.
The embedding of this hypersurface can be described in terms of
$N_V$ complex coordinates $z^A$ 
($A=1,\dots ,N_V$) by letting the $X^I$ be proportional to
some holomorphic sections $X^I(z)$ of the complex projective space.
In terms of these 
sections the $X^I$ read
\beqa
X^I = e^{{1\over 2}K(z,\bar z)}\,
X^I(z),
\eeqa
where $K(z,\bar z)$ is the K\"ahler potential, to be introduced 
below. 
The resulting geometry for the space of physical scalar fields
belonging to vector multiplets of an $N=2$ 
supergravity theory is a special K\"ahler geometry,
with a K\"ahler metric $g_{A\bar B}=\partial_A\partial_{\bar B}K(z,\bar z)$,
being the metric of the moduli space ${\cal M}_V$,
following from a K\"ahler potential of the special 
form
\begin{equation}
K(z,\bar z)=
-\log\Big(i \bar X^I(\bar z)F_I ( X^I(z)) -i X^I(z)
\bar F_I(\bar X^I(\bar z))\Big) .
\label{KP} 
\end{equation}
The holomorphic sections transform under 
projective transformations $X^I(z)\to e^{f(z)}X^I(z)$, which 
induce a K\"ahler transformation on the K\"ahler potential $K$ and a 
U(1) transformation on the section $V$, 
\beqa
K(z,\bar z)\rightarrow K(z,\bar z)-f(z)-\bar f(\bar z)\,, \qquad 
V(z,\bar z) \rightarrow {\rm e}^{{1\over 2}(f(z)-\bar f(\bar 
z))}V(z,\bar z)\,.   \label{kweight}
\eeqa
A convenient choice of inhomogeneous coordinates $z^A$
are the {\em special} coordinates, defined by 
$
X^0(z)=1$, $X^A(z)= z^A$, $A=1,\ldots ,N_V$.
In this parameterization the K\"ahler potential can be written as 
\begin{equation}
K(z,\bar z) = -\log\Big(2({\cal F}+ \bar{\cal F})-
           (z^A-\bar z^A)({\cal F}_A-\bar{\cal F}_A)\Big)\,,
\label{Kspecial}
\end{equation}
where ${\cal F}(z)=i(X^0)^{-2}F(X)$.

Besides $V$, the magnetic/electric field strengths 
$(F_{\mu\nu}^I, G_{\mu\nu I})$ also consitute a symplectic 
vector. Here $G_{\mu\nu I}$ is generally defined by $G^+_{\mu\nu
I}(x) = -4i g^{-1/2}\, \d S/\d F^{+\m\n I}$. 
Consequently, also the corresponding magnetic/electric charges 
$Q=(p^I, q_I)$ transform as a symplectic vector.
The Lagrangian terms containing the kinetic energies of
the gauge fields are
\begin{equation}
4 \pi  {\cal L}^{\rm gauge} = -{\textstyle{i\over 8}}
( {\cal N}_{IJ}\,F_{\mu\nu}^{+I}F^{+\mu\nu J}
- \bar{\cal N}_{IJ}\,F_{\mu\nu}^{-I} F^{-\mu\nu J} ),\label{n2gaugekin}
\end{equation}
where $F^{\pm I}_{\mu\nu}$ denote the selfdual and anti-selfdual 
field-strength components and the field dependend
${\cal N}_{IJ}(z,\bar z)$, which are essentially given by the second 
derivative of the prepotential,  
comprises the inverse gauge couplings
$g^{-2}_{IJ}= {i\over 16\pi}({\cal N}_{IJ}-\bar {\cal N}_{IJ})$
and the generalized $\theta$ angles
$\theta_{IJ}= \frac{\pi}{2} ({\cal N}_{IJ}+\bar {\cal N}_{IJ})$.

\subsection{The Heterotic Prepotential}

Now, we will discuss the
 heterotic 
$N=2$ models, obtained by compactifying
the $E_8 \times E_8$ string on $K3\times T_2$.  The  moduli $z^A$
($A=1,\dots, N_V$)
comprise the dilaton $S$, the two toroidal moduli $T$ and $U$
as well as Wilson lines $V^i$ ($i=1,\dots ,N_V-3$):
\begin{equation}
S= -i z^1, T= -i z^2 ,
U= -i z^3 , V^i= -i z^{i+3}.
\label{modid}
\end{equation}
We will, in the following, 
collectively denote the moduli $T,U$ and $V^i$ by $T^a$, so that  
$a=2,\ldots,N_V$.
For this class of models, the heterotic prepotential has the  form
\beqa
{\cal F}^{\rm het}= 
- S T^a\eta_{ab}T^b  + h(T^a)  + 
f^{\rm NP}(e^{- 2\pi S},T^a) \;,
\label{hetprepot}
\eeqa
where
$
T^a\eta_{ab}T^b=T^2T^3-\sum_{I=4}^{N_V}(T^I)^2$.
The first term in (\ref{hetprepot}) is the classical part of the 
heterotic prepotential, $h(T^a)$ denotes the one-loop contribution
and $f^{\rm NP}$ is the non-perturbative part, which is 
exponentially suppressed for small coupling. 

The classical prepotential leads to the metric of the special
K\"ahler manifold eq.(\ref{mv}) with corresponding tree-level K\"ahler potential
\begin{equation}
K       = -\log \Bigl[(S + \bar S)\Bigr ] -
        \log\Bigl[ (T^a + \bar T^a)\eta_{ab}(T^b + \bar T^b)\Bigr].
\label{Ktree}
\end{equation}
Note that the classical 
`Yukawa' couplings ${\cal F}_{ABC}={\partial {\cal F}\over \partial z^A
\partial z^B\partial z^C}$ have precisely the same
form as the Calabi-Yau intersection
numbers $C_{ABC}$ in the limit $t_S\rightarrow\infty$ 
(see eq.(\ref{intersecno})).
Remember that this matching of `Yukawa' couplings in the weak coupling
limit was one of the reasons to consider $K3$ fibrations on the dual type II
side.

Due to the required embedding of the $T$-duality group
into the $N=2$ symplectic transformations, it follows \cite{DKLL,AFGNT} 
that the heterotic one-loop prepotential $h(T^a)$ must 
obey well-defined transformation rules under this
group.
The function $h(T^a)$ leads to the following modified 
K\"ahler potential \cite{DKLL}, which represents the full 
perturbative contribution,
\begin{eqnarray}
K = -\log\lbrack (S +  \bar S )+V_{GS}(T^a , \bar T^a)\rbrack  -
\log\Bigl[  (T^a + \bar T^a)\eta_{ab}(T^b + \bar T^b) \Bigr] ,
\label{Kloop}
\end{eqnarray}
where
\begin{eqnarray}
 V_{GS}
= \Biggl(2( h + \bar h) - (T^a + \bar T^a)(\partial_{T^a} h 
 +  
\partial_{\bar T^a}
\bar h)\Biggr)
/ (T^a + \bar T^a)\eta_{ab}(T^b + \bar T^b)
\label{GSfunction}
\end{eqnarray}
is the Green-Schwarz term  describing the mixing
of the dilaton with the moduli $T^a$. Due to this mixing,
the $S$-field is not any more invariant under the target-space duality 
transformations at the one  loop level. 
However it is possible to define a modular invariant 
dilaton field by simply shifting $S$ by a function of the other moduli like
\begin{equation}
S_{\rm inv}=S+{1\over 2(N_V+6)}\lbrack i\eta^{ab}\partial_a\partial_bh+L\rbrack,
\end{equation}
where $L$ is a modular invariant holomorphic function given below.

Let us now discuss the precise form of the one-loop
prepotential $h(T^a)$
\cite{HM,CCLM,CCL} for some specific heterotic
compactifications on $K3\times T^2$. 
Namely we will consider the models with instanton embedding
$k=0,1,2$, and will first start with those models
with $N_V=4$ and $N_H=215-12k$.
The four vector multiplets are denoted by $S$, $T$, $U$ and the
Wilson line $V$. Later we
will discuss the Higgs transition via an enhanced $SU(2)$ gauge group
to the $S-T-U$ models with
$N_V=3$ and $N_H=244$.
For the class of $S$-$T$-$U$-$V$ models considered
here, the one-loop prepotential is given by
\begin{eqnarray}
h= p_n(T,U,V) 
- c  -  \frac{1}{4\pi^3}\sum_{n,l,b\in {\bf Z} \atop
(n,l,b)>0} c_k(4nl-b^2)   Li_3({\bf e}[inT+ilU+ibV]), 
\label{prepstuv}
\end{eqnarray}
where
$c=\frac{c_{n}(0) \zeta(3)}{8 \pi^{3}}$ and 
${\bf e}[x]={\rm exp} 2 \pi i x$.
The coefficients $c_k(4nl-b^2)$ are the expansion coefficients of 
particular Jacobi modular forms \cite{CCL}\footnote{A Siegel 
modular form $F(T,U,V)$, which is an authormorphic function
of the discrete group $Sp(4,{\bf Z})$ of weight $n$,
has a Fourier expansion with respect to its variable $iU$,
$
F(T,U,V)=\sum_{m=0}^{\infty}\phi_{n,m}(T,V)s^m 
$,
where $s={\bf e}[iU], {\bf e}[x] = {\rm exp}2 \pi i x$.
Each of the $\phi_{n,m}(T,V)$ is a Jacobi form
of weight $n$ and index $m$.  
A Jacobi form $\phi_{n,m}(T,V)$ of index $m$ has 
in turn 
an
expansion
$
\phi(T,V)=\sum_{n\ge 0}\sum_{l\epsilon {\bf Z}}c(n,l)q^nr^l 
$,
where $q={\bf e}[iT], r = {\bf e}[iV]$.
Consider, for instance, the Eisenstein series, which have 
the expansion
$
{\cal E}_n(T,U,V) = E_n(T) - \frac{2n}{B_n}\, E_{n,1}(T,V)\, s + {\cal O}(s^2)
$.
The explicit expansion coefficients of $E_{4,1}$ and $E_{6,1}$ are given
in \cite{CCL}. In addition,
a review of several 
properties of Siegel and Jacobi modular forms is given in the appendix
of \cite{CCL}}:
\begin{eqnarray}
  {1\over \Delta(\tau)}
\Biggl(\frac{12+k}{24}E_6(\tau) E_{4,1}(\tau,z)  +
E_4(\tau)\frac{12-k}{24}E_{6,1}(\tau,z)\Biggr) =
\sum_{n,b}c_k(4n-b^2)q^nr^b,\label{e41jac}
\end{eqnarray}
$\Delta=\eta^{24}(\tau)$, $q=e^{2\pi i\tau}$, $r=e^{2\pi i z}$.
This expression is very closely related to the elliptic genus of $K3$,
$\frac{1}{\eta^2}Tr_R F(-1)^F q^{L_0-c/24}\bar{q}^{\tilde{L}_0-\tilde{c}/24}$,
with gauge bundle characterized by the integer $k$; it can be explicitly
computed in string perturbation theory.

\noindent $p_k$ is a cubic polynomial of the form \cite{HM,CCL}
\begin{eqnarray}
p_k(T,U,V)=-{\textstyle{1\over 3}}U^3-({\textstyle{4\over 
3}}+k)V^3+(1+{\textstyle{1\over 2}}k)UV^2+{\textstyle{1\over 2}}k
TV^2 \;\;.\label{cubicfa}
\end{eqnarray}

The (Wilsonian) 
Abelian gauge threshold functions are related 
to the second derivatives
of the one-loop prepotential. 
At the loci of enhanced non-Abelian gauge symmetries some of the
Abelian
gauge couplings will exhibit logarithmic singularities due
to the additional massless states.
First consider $\partial_T\partial_U h$.
At the line $T=U$  one $U(1)$ is extended to $SU(2)$ without
additional massless hypermultiplets. 
It can  be easily checked that, as $T \rightarrow U$,
\beqa
\partial_T \partial_U h = -\frac{1}{\pi} \log(T-U) \;\;,\label{logtu}
\eeqa
as it should.
The Siegel modular form which vanishes on the $T=U$ locus and has
modular weight $0$ is given by $\frac{{\cal C}^2_{30}}
{{\cal C}_{12}^5}$.
It can be shown that, as $V \rightarrow 0$,
\beqa
 \frac{{\cal C}^2_{30}}{{\cal C}_{12}^5} \rightarrow (j(T) - j(U))^2\;\;,
 \label{siegel30}
\eeqa
up to a normalization constant.
Hence one deduces that
\beqa
\partial_T \partial_U h =
-\frac{1}{2\pi} \log \frac{{\cal C}^2_{30}}{{\cal C}_{12}^5}+ {\rm regular}.
\label{tugaugecoupl}
\eeqa
On the other hand, at the locus $V=0$, a different $U(1)$ gets enhanced to
$SU(2)$, and at the same time $N_H'=12k+32$ hyper multiplets, 
being doublets of $SU(2)$,
become massless. 
The Siegel modular form which vanishes on the $V=0$ locus and has
modular weight $0$ is given by $\frac{{\cal C}_5}
{{\cal C}_{12}^{5/12}}$, and can derive that
\beqa
-{1\over 4}\partial_V^2 h = \frac{3}{4\pi} (2 + k) 
\log \left(\frac{{\cal C}_5}
{{\cal C}_{12}^{5/12}}\right)^2+ {\rm regular}\label{siegel5}
\eeqa
The coefficient of the $\log$ in this equation is just the $N=2$ 
$\beta$-function coefficient of $SU(2)$ with $12k+32$ massless doublets.

Let us briefly investigate the Higgs transition to the $S$-$T$-$U$
models with $N_V=3$ and $N_H=244$. This transition takes place at the locus
$V=0$ giving vev's to the $N_H'$ new massless hypermultiplets (3 of them
will eaten by the $SU(2)$ gauge bosons).
It follows that the 
effective action of the $S$-$T$-$U$ models can be obtained from the 
$S$-$T$-$U$-$V$
prepotential by setting $V=0$ in eq.(\ref{prepstuv}).
Then, the sum over $b$
in (\ref{prepstuv}) yields independently from $k$ the coefficients of the 
$3$ parameter model,
\beqa
c_{STU}(nl)=\sum_b c_k(4nl-b^2)  \;\;.\label{cstu}
\eeqa
Via the properties of the Jacobi functions, we get that
\begin{eqnarray}
\frac{1}{\Delta}E_4(\tau)E_6 (\tau)
=\sum_{n\ge -1} c_{STU}(n)q^n
\;\;.\label{e4mod}
\end{eqnarray}
This result can be indepedently checked
by computing $c_{STU}(n)$ directly in the $S$-$T$-$U$ model \cite{CCLM}.

The one-loop gauge couplings of the $S$-$T$-$U$ models
exhibit logarithimic singularities at the locus $T=U$ \cite{CLM2,DKLL,FOST}:
\begin{equation}
\partial_T\partial_Uh=-{1\over \pi}\log (j(T)-j(U))+{\rm regular}.\label{logj}
\end{equation}
Moreover the modular transformation properties of $h$ allow us
to uniquely determine the Yukawa couplings \cite{DKLL,AFGNT}; e.g.
\begin{equation}
\partial^3_Th={2\over \pi}{E_4(T)E_4(U)E_6(U)\over \eta^{24}(U)(j(T)-j(U))}.
\label{loopyuk}
\end{equation}
By performing the $q$-expansion of this expression, this
equation can be checked
by taking the third derivative of the prepotential eq.(\ref{prepstuv}), 
and also expanding it up to some order in $q$.


\subsection{The Type-IIA Prepotential and the Heterotic/type IIA Map}

As already mentioned, the prepotential in type-IIA Calabi--Yau 
compactifications, which depends on 
 the K\"ahler-class moduli $t_A=-iz^A$ ($A=1,\dots ,N_V=h_{1,1}$),
is of purely classical origin. 
The type-IIA prepotential has the following general structure 
\cite{HKTY,hosono}: 
\begin{eqnarray}
 {\cal F}^{\rm II} = -{1\over 6}C_{ABC}t_At_Bt_C 
- \frac{\chi \zeta(3)}{2 (2 \pi^{3})}+
\frac{1}{(2 \pi)^3} \sum_{d_1,...,d_h}
n^r_{d_1,...,d_h} Li_3({\bf e} [i\sum_Ad_At_A]),
\label{ftype2}
\end{eqnarray}
where we work inside the K\"ahler cone ${\rm Re}t_A\geq 0$.
The cubic part of the type-IIA
prepotential is given in terms of the classical intersection
numbers $C_{ABC}$ and the Euler number $\chi$, 
whereas the coefficients $n^r_{d_1,...,d_h}$
of the exponential terms denote the rational instanton numbers of 
genus 0 and multi degree $d_A$.
They can be computed
by the mirror map studying
the corresponding type IIB compactification on the
mirror Calabi-Yau space.
The rational instantons are related to non-perturbative
effects on the world-sheet, namely they
count the  embeddings of the genus
0 world-sheet into the Calabi-Yau space; their
contributions disappear in large radius limit
$\alpha'/R\rightarrow 0$. 
Hence, in the limit of large K\"ahler class moduli, 
$t_A\rightarrow\infty$, 
only the classical part, related to the intersection numbers, survives.

To be specific let us consider those three Calabi-Yau space with $h^{(1,1)}=4$
and $h^{(2,1)}=214-12k$, $k=0,1,2$. Recall that the $k=2$ Calabi-Yau
can be constructed as hypersurface in $WP_4^{(1,1,2,6,10)}(20)$.
The classical intersection numbers are given in \cite{LSTY,BKKM}, 
and the cubic part
of the type II prepotentials for these three models can be written as follows
\beqa
-{\cal F}^{II}_{\rm cubic}&=&t_1(t_2^2+t_2t_3+4t_2t_4+2t_3t_4+3t_4^2)
+{4\over 3}t_2^3+8t_2^2t_4+{k\over 2}t_2t_3^2
+(1+{k\over 2})t_2^2t_3 \nonumber \\ &+&2(k+2)t_2t_3t_4
+kt_3^2t_4
+(14-k)t_2t_4^2 
+ (4+k)t_3t_4^2+(8-k)t_4^3 . \label{cubicf}
\eeqa
In order to match (\ref{cubicf})
with the cubic part of the heterotic prepotential given in 
(\ref{prepstuv}), we have to perform the following identification of
type II and heterotic moduli 
\beqa
 t_1 &=& S-{k\over 2}T-(1-{k\over 2})U,\qquad t_2=U-2V,\nonumber\\
t_3&=&T-U,\qquad t_4=V  \;\;,
\label{coordin}
\eeqa
Since we are working inside the K\"ahler cone ${\rm Re}t_A\geq 0$,
the heterotic variables $S$, $T$, $U$ and $V$ have to obeye the following
inequalities:
\begin{equation}
 T\geq U\geq 2V\geq0,\quad  S\geq {k\over 2}T-(1-{k\over 2})U.\label{ordering}
\end{equation}

Next, let us consider the contributions of the world sheet 
instantons to the type II prepotential of  the 4 parameter models.
The $n^r_{d_1,d_2,d_3,d_4}$ denote the rational instanton numbers.
The heterotic weak coupling limit $S \rightarrow \infty$ corresponds
to the large K\"ahler class limit $t_1 \rightarrow \infty$.  In this
limit, only the instanton numbers with $d_1=0$ contribute
in the above sum.  Using the identification 
$nT+lU+bV=d_1t_1+d_3t_3+d_4t_4$, it follows that 
$
n=d_3$, 
$l=d_2-d_3$,   
$b=d_4-2d_2$
Then, the instanton part of the prepotential turns into
\beqa
{\cal F}^{II}_{\rm inst} ={1\over (2\pi)^3} \sum_{n,l,b}
n^r_{0,l+n,n,b+2l+2n}
Li_3(e^{-2\pi(nT+lU+bV)}).
\label{largespr}
\eeqa
Matching this expression with eq.(\ref{prepstuv}) 
shows that the rational instanton
numbers have to satisfy the following constraint
\beqa
n^r_{0,l+n,n,b+2l+2n} = - 2 c_k(4nl-b^2) \;\;.
\label{insmod}
\eeqa
This relation is very non-trivial and connects the rational instanton numbers
to the coefficients of the modular functions which determine the heterotic
prepotential.
For concreteness,  above relation (\ref{insmod}) can be successfully checked
for the $k=2$ model, which is  based on the
Calabi-Yau space $WP_4^{(1,1,2,6,10)}(20)$, 
using the instanton numbers given in \cite{BKKM}.

Let us discuss some properties of the heterotic/type IIA map
 in a slightly simpler
situation, namely in the $S$-$T$-$U$ models; the Higgs transition from 
$h^{(1,1)
}=4$ to the $h^{(1,1)}=3$ models takes place at the boundary of the
K\"ahler cone $t_4=0$; the cubic part of the prepotential of the
$S$-$T$-$U$ models  is then simply obtained
by setting $t_4=0$ in eqs.(\ref{cubicf}) and (\ref{coordin}).
In addition, the instanton numbers 
$n^{r}_{d_1,d_2,d_3}$ of the $S$-$T$-$U$ model are 
given by \cite{BKKM}
\beqa
n^{r}_{d_1,d_2,d_3} = 
\sum_{b} n^r_{d_1,d_2,d_3,b} \;\;,\label{truncsum}
\eeqa
where the summation range over $b$ is finite.  
In order that the heterotic and the type IIA prepotentials match in the weak
coupling limit $S,t_1\rightarrow\infty$,
the instanton numbers $n^r_{0,l+n,n}$ must be related to the coefficients
$c_{STU}(n)$
of the modular form $E_4E_6/\Delta$ in the following way \cite{HM,CCLM}
\begin{equation}
n^r_{0,l+n,n}=-2c_{STU}(ln).\label{stunr}
\end{equation}
Again one can check by direct computation of some instanton numbers
that this relation is satisfied for the $k=2$ model based on the Calabi-Yau
space $WP_4^{(1,1,2,8,12)}(24)$. In addition one can show 
\cite{KV,KLT} for this model
that the Calabi-Yau `Yukawa couplings' in the weak coupling limit agree
with the heterotic Yukawa couplings eq.(\ref{loopyuk}), as predicted from the
$K3$ fibration of this Calabi-Yau.

By  inspection of the instanton numbers the quantum symmetries of the
$S$-$T$-$U$ vacua become manifest. Specifically, for the $k=2$ model,
two symmetries  among the instanton numbers are discovered \cite{klm}, 
namely $n^r_{0,d_2,d_3}=n^r_{0,d_2,d_2-d_3}$ and $n^r_{d_1,d_2,d_3}=
n^r_{d_3-d_1,d_2,d_3}$.
So the theory is invariant under the following two shifts
\begin{eqnarray}
&~ & t_2\rightarrow t_2+t_3,t_3\rightarrow -t_3,\quad{\rm for}\quad t_1=\infty
\nonumber\\
&~& t_1\rightarrow -t_1, t_3\rightarrow t_1+t_3.\label{qsym}
\end{eqnarray}
Using the identification $t_1=S-T$, $t_2=U$ and $t_3=T-U$,
the first symmetry corresponds to the perturbative exchange
symmetry $T\leftrightarrow U$ for $S\rightarrow\infty$, whereas the second
is the non-perturbative exchange symmetry $S\leftrightarrow T$.
The line $S=T$ is the border of the K\"ahler cone, ${\rm Re}t_1\geq 0$,
of the $k=2$ model and is the locus of the strong coupling singularity.

Furthermore comparing the intersection numbers (see eq.(\ref{cubicf}))
and the instanton numbers of the
$k=0$ and $k=2$ $S$-$T$-$U$ models it can be seen that the prepotentials
of these two models become equivalent upon the substitution
$t_1\rightarrow t_1+t_3$. In fact, the $k=0$ model is completely invariant
under the exchange of $t_1$ and $t_3$, which corresponds to an
exchange of the two $P^1$'s which serve as the base of the two alternative
$K3$ fibrations. In the $k=0$ model the identfication among the heterotic
and type IIA variables is given as $t_1=S-U$, $t_2=U$ and $t_3=T-U$. This shows
that the K\"ahler cone of the $k=0$ model contains both regions
$S>T$ and $S<T$ in agreement with the absence of the strong coupling
singularity along the line $S=T$. It means that the K\"ahler cone of
the $k=2$ model is a subcone of the K\"ahler cone of the $k=0$ model.
Considering also the $k=1$ model, there would be also a linear
transformation, $t_1\rightarrow t_1+{1\over 2}t_3$, which maps the intersection
numbers of the $k=1$ model and the $k=2$ model into each other. However,
since this transformation contains half-integer coefficients,
the instanton parts of the prepotentials do not agree. Therefore the
$k=1$ vacumm is physically different from the $k=0,2$ vacua after including
the world-sheet instanton effects.


\subsection{The gravitational coupling $F^{(1)}$}

The quantitative tests of the heterotic/type II string-string duality
can be continued by considering higher order couplings in the effective
$N=2$ supergravity action.
The higher derivative couplings of vector multiplets $X$ to the Weyl multiplet
$W$ can be expressed as a power series
\beqa
F(X,W^2) = \sum_{g=0}^\infty F^{(g)}(X) \, \big[W^2\big]^g
\,. \label{holoexpansion}
\eeqa
$F^{(0)}$ is just the prepotential discussed before.
We will restrict the discussion to the function $F^{(1)}$ which appears as
the field dependent coupling of the gravitational term $R^2$, $R$ being
 the Riemmann tensor, in the effective action.
Higher order terms and a recursion between terms of different order in $g$
were discussed in \cite{AGNT,dewit,DCLMR}. For simplicity, we will  
restrict
the discussion to the $S$-$T$-$U$-models with  $k=0,1,2$. Explicit results
on the $S$-$T$-$U$-$V$ models can be found in \cite{CCL}.

The heterotic perturbative 
(tree level plus one loop) Wilsonian gravitational coupling 
for the $S$-$T$-$U$ model
is given by \cite{CLM2,KLT}
\beqa
 F^{(1)} = 24 S_{\rm inv} 
- \frac{b_{\rm grav}}{\pi} \log \eta(T) \eta(U)
 + \frac{2}{\pi} \log (j(T)-j(U)),
\label{f1stu}
\eeqa
where \cite{DKLL}
$
S_{\rm inv} =
S - \sigma(T,U)$, $\sigma(T,U)=-{1\over 2}\partial_T\partial_U h+{1\over 8}
L(T,U)$, $L(T,U)=-{4\over \pi}\log(j(T)-j(U))$.
$F^{(1)}$ is singular at the line $T=U$ due to the additional massless
states at $T=U$. The term proportional $b_{\rm grav}$ is needed in order
to cancel the gravitational anomaly due to the massless states.
Using some product formula for $j(T)-j(U)$, 
the function $ F^{(1)}$ can be written \cite{CCLM}
as a power series expansion with coefficients $\tilde c_1$, where
where the coefficients $\tilde{c}_1$ are given by \cite{HM}
$
\frac{E_2(\tau)E_4(\tau)E_6(\tau)}{\Delta(\tau)} = \sum \tilde{c}_1(n) q^n 
$.

In the dual type IIA string, the gravitational coupling $ F^{(1)}$
arises at the one-loop level (genus one), and is determined by the rational
instanton numbers $n^r$ as well as by the elliptic instanton numbers
$n^e$:
\begin{eqnarray}
F^{(1)}=-i\sum_{A=1}^{h^{(1,1)}}t_Ac_{2A}
-{1\over \pi}\sum_n\Biggl(
12n^e_{d_1,\dots ,d_h}\log(\tilde\eta(\prod_{A=1}^{h^{(1,1)}}q_A^{d_A}))
+n^r_{d_1,\dots ,d_h}\log(1-\prod_{A=1}^{h^{(1,1)}}q_A^{d_A})\Biggr).
\label{f1typeII}
\end{eqnarray}
Here $\tilde\eta(q)=\prod_{m=1}^\infty(1-q^m)$, 
the $c_{2A}$ are the 2nd Chern
numbers of the Calabi-Yau space and the $n^e_{d_1,\dots ,d_A}$
are the elliptic instanton numbers which count elliptic genus one curves
embedded into the Calabi-Yau space.
In the weak coupling limit $S\rightarrow\infty$, the heterotic
and the type IIA  $F^{(1)}$ functions must agree. By this requirement
one can derive \cite {CCLM} the following interesting relation
among rational, elliptic instanton numbers and $\tilde c_1$ coefficients:
\beqa
12\sum_{i=1}^sn^e_{0,d_2^i,d_3^i}+n^r_{0,l+k,k}=-2\tilde c_1(kl),\label{ellins}
\eeqa
where $s$ is the number of common divisors $m_i$ ($i=1,\dots , s$)
of $d_2=k+l$ and $d_3=k$ with $d_{2,3}^i=d_{2,3}/m_i$ (where $m_2=1$).
Considering again the 3 parameter Calabi-Yau $WP_4^{(1,1,2,8,12)}(24)$
one can explicitly compute some of the elliptic instanton numbers and check
in this way that the relation (\ref{ellins}) indeed holds.

\section{$N=2$ Black Hole Solutions and their Entropies}

One of the celebrated successes within the recent non-perturbative
understanding of string theory and M-theory is the matching of the
thermodynamic Bekenstein-Hawking black-hole entropy with the
microscopic entropy based on the counting of the relevant D brane
configurations which carry the same charges as the black hole
\cite{StromVafa}.
 This comparison works nicely for
type-II string, respectively, M-theory backgrounds which break half of the
supersymmetries, i.e.  exhibit $N=4$ supersymmetry in four
dimensions.
For example, consider the type IIA superstring compactified on $K3\times T^2$.
Then, the intersection of three D 4-branes, whose spatial parts of their
world volumes are wrapped $p^A$ times ($A=1,2,3$)
around the
internal 4-cycles, together with
$q_0$ D 0-branes leads to a four-dimensional black hole with electric
charge $q_0$ and magnetic charges $p^A$. This black hole has non-vanishing
event horizon $A$, and the corresponding Bekenstein-Hawking entropy
is given by the following expression:
\begin{equation}
{\cal S}_{\rm BH}= {A\over 4}=2\pi\sqrt{q_0p^1p^2p^3} . \label{area}
\end{equation}

Here we will discuss black hole solutions and
corrections to this formula which arise
in string backgrounds with $N=2$ supersymmetry in four
dimensions.
In particular, studying IIA compactifications
on a (complex) 3-dimensional Calabi-Yau space,
corrections arise due to the internal geometry and topology of the 
Calabi-Yau manifold \cite{BCDWKLM}. In the large volume limit
of the Calabi-Yay space, the Calabi-Yau intersection numbers enter
the entropy formula, since the D 4-branes are now wrapped around
the non-trivial Calabi-Yau 4-cycles. Furthermore, in case of generic radii
also world-sheet instanton corrections, encoded by the rational
instanton numbers determine the black hole solutions and their
entropies. 
In addition, we will consider corrections to the $N=2$ black hole entropies
which arise from higher-derivative terms involving higher-order products of 
the Riemann tensor and the vector field strengths \cite{msw,higher}.

From the point of view of the low energy effective lagrangian the
wrapped D branes  correspond to extremal charged black hole
solutions of the effective $N=2$ supergravity.
Extremal, charged, $N=2$ black holes, their entropies and also the
corresponding brane configurations were discussed in several recent
papers 
\cite{FeKaStr,n2solut,BCDWKLM,BehrndtLuestSabra,BehLu,msw,higher}. 
 One of 
the key features of extremal $N=2$ black-hole solutions
is that the moduli depend in general on $r$, but show a fixed-point
behaviour at the horizon. This fixed-point behaviour is implied 
by the fact that, at the  
horizon, full $N=2$ supersymmetry is restored; at the 
horizon the metric is equal to the Bertotti-Robinson 
metric, corresponding to the $AdS_2\times S^2$ geometry. 
The extremal black hole can be 
regarded as a soliton solution which interpolates between
two fully $N=2$ supersymmetric vacua, namely corresponding to $AdS_2\times 
S^2$ at the horizon and flat Minkowski spacetime at spatial 
infinity.

\subsection{Extremal black holes in $N=2$ supergravity}

Now we want to discuss the static $N=2$ black hole solutions of
the lowest order $N=2$ supergravity action which we introduced in
section (3.1). In particular we have seen that in the context of special
geometry the two fundamental objects of the $N=2$ supergravity action are
given by  two symplectic vectors,
namely the period vector $V=(X^I,F_I)$ and the magnetic/electric field
strength vector
$(F_{\mu\nu}^I, G_{\mu\nu I})$ respectively by $V=(X^I,F_I)$ and
the magnetic/electric charge vector $Q=(p^I, q_I)$.

In terms of 
these symplectic vectors the stationary solutions have been 
discussed in \cite{BehrndtLuestSabra,higher}. 
Here one solved the generalized  
Maxwell equations in terms of  
$2N_V+2$ harmonic functions, which therefore also transform as a 
symplectic vector ($m,n=1,2,3$), 
\begin{equation}
F_{mn}^I={1\over 2}\epsilon_{mnp}\,\partial_p\tilde H^I(r)\quad ,
\quad G_{mn I}={1\over 2}\epsilon_{mnp}\,\partial_pH_I(r)\, .\label{gaugeh}
\end{equation}
The harmonic functions can be parametrized as
\begin{equation}
\tilde H^I(r)  = \tilde h^I+{p^I\over r}\,, \qquad 
H_I(r)=h_I+{q_I\over r}\,,  \label{harmonic}
\end{equation}
and we write the corresponding symplectic vector as $H(r)=(\tilde 
H^I(r), H_I(r)) = h + Q/r$. 

Next we want to find the solutions for the four-dimensional black hole metric
and for the scalar moduli fields
$z^A=X^A/X^0$, which break half of the $N=2$
supersymmetries.
For the metric we make the ansatz
\cite{tod}
\begin{equation}
ds^2 = - e^{2U} dt^2 + e^{-2 U} dx^m dx^m \ ,\label{metrican}
\end{equation}
where $U$ is a function of the radial coordinate $r=\sqrt{x^m x^m}$. 
As shown in \cite{BehrndtLuestSabra} the Killing spinor
equations follow from the symplectic covariance of the vectors
$V$ and $H$, namely 
these vectors should satisfy a certain proportionality relation. The 
simplest possibility is to assume that $V$ and $H$ are directly 
proportional to each other. Because $H$ is real and 
invariant under U(1) transformations, there is a complex 
proportionality factor, which we denote by $Z$. Hence we define
a U(1)-invariant symplectic vector (here we use the homogeneity 
property of the function $F$), 
\begin{equation}
\Pi=\bar Z V=(Y^I,  F_I(Y)) \,,\label{newsec}
\end{equation}
so that $Y^I=\bar ZX^I$, and set 
\begin{equation}
\Pi (r) -\bar \Pi(r) = i H(r)\,. \label{stable}
\end{equation}
The solutions of this set of algebraic equations, which we
call stabilization equations and which depend on the
particular choice of the prepotential, fully determine the form of the
extremal BPS
black holes. Specifically, 
the solution for the black hole metric is given by the symplectically
invariant ansatz
\begin{equation}
e^{-2U(r)} = Z(r) \bar Z(r),
\end{equation}
where $Z(r)$ is determined as
\begin{equation}
Z(r)= - H_I(r)\, X^I + \tilde H^I(r)\, F_I(X)\,,\qquad \vert 
Z(r)\vert^2 = i \langle 
\bar \Pi(r) ,\Pi(r)\rangle \,.  \label{Z}
\end{equation}
The stabilization equations equations (\ref{stable})
also determine the $r$-dependence of
the scalar moduli fields: 
\begin{equation}
z^A(r)=Y^A(r)/ Y^0(r).
\end{equation}
 So the constants
$(\tilde h^I,h_I)$ just determine the asymptotic values of the scalars at
$r=\infty$. In order to obtain an asymptotically flat metric with 
standard normalization, these
constants must fullfill some constraints. Near the horizon 
($r\approx 0$), (\ref{stable})
takes the form used in \cite{BCDWKLM} and $Z$ becomes proportional 
to the holomorphic BPS mass ${\cal M}(z) = q_I X^I(z) - p^I F_I (X(z))$.
It can  
be shown that the solution preserves half the supersymmetries, 
except at the horizon and at spatial infinity, where 
supersymmetry is unbroken.

{From} the form of the static solution 
at the horizon ($r\rightarrow 0$) we can easily derive its macroscopic
entropy. Specifically the Bekenstein-Hawking entropy is given by 
\begin{eqnarray}
{\cal S}_{\rm BH}&=&\pi \,(r^2e^{-2U})_{r=0}=\pi\,(r^2Z\bar Z)_{r=0}
\nonumber\\
&=&i\pi\,\Big(\bar Y^I_{\rm hor}\,F_I(Y_{\rm hor})-\bar F_I
(\bar Y_{\rm hor})\,Y^I_{\rm hor}\Big)\,,\label{entropy}
\end{eqnarray}
where the symplectic vector $\Pi$ at the horizon,
\begin{equation}
\Pi(r)\stackrel{r\to 0}{\approx} {\Pi_{\rm hor}\over r}\,,\qquad 
Y^I(r)\stackrel{r\to 0}{\approx} {Y^I_{\rm hor}\over r} \ ,\label{qhor}
\end{equation}
is determined by the following set of stabilization equations:
\begin{equation}
\Pi_{\rm hor}-\bar\Pi_{\rm hor} = i Q.\label{stabhor}
\end{equation}
So we see that the entropy as well as the scalar fields $z^A_{\rm hor}=
Y^A_{\rm hor}/ Y^0_{\rm hor}$ depend only on the magnetic/electric
charges $(p^I,q_I)$. It is useful to note that 
the set of stabilization equations (\ref{stabhor})
is equivalent to the minimization of $Z$ with respect to
the moduli fields \cite{FeKaStr}.

\subsection{$N=2$, Type II Calabi-Yau superstring vacua}

\subsubsection{The large radius limit of type IIA Calabi-Yau compactifications}

As an example, consider a type-IIA compactification on a Calabi-Yau
3-fold. The number of vector superfields is given as $N_V =h^{(1,1)}$.
As discussed
in section (3.3) the prepotential, which is purely classical, contains the
Calabi-Yau intersection numbers of the 4-cycles, $C_{ABC}$, 
and, as $\alpha'$-corrections, the Euler
number $\chi$\footnote{The corresponding term
in the effective supergravity action
comes from an higher derivative $R^4$ term
in ten dimensions.}
and the rational instanton numbers $n^r$. Hence the
black-hole solutions will depend in general on all these topological
quantities \cite{BCDWKLM}. 
However, for a large Calabi-Yau volume, i.e. large values of the K\"ahler
class moduli fields $z^A=X^A/X^0$,
only the
part from the intersection numbers survives and the $N=2$ prepotential
is given by 
\begin{equation}
F(Y)=D_{ABC}{Y^AY^BY^C\over Y^0},\qquad D_{ABC}=- {1\over 6} C_{ABC}.
\label{cprep}
\end{equation}
Based on this prepotential we consider in the following 
a class of non-axionic
black-hole solutions (that is, solutions with purely imaginary moduli 
fields $z^A$, i.e. real moduli fields $t_A$) 
with only
non-vanishing charges $q_0$ and $p^A$ ($A=1,\dots ,h^{(1,1)}$). So
only the harmonic functions $H_0(r)$ and $\tilde H^A(r)$ are nonvanishing.
This charged configuration corresponds, in the type-IIA 
compactification, 
to the intersection of three D4-branes, wrapped over the internal Calabi-Yau 
4-cycles and hence carrying magnetic charges $p^A$, plus one D0-brane
with electric charge $q_0$. 
For the configuration indicated above, the solutions of the
stabilization equations
eq.(\ref{stable})
have the following form:
\begin{equation} 
Y^0(r) = {\lambda (r)\over 2} \qquad , \qquad Y^A(r) = 
i {\tilde H^A (r)\over 2}
\end{equation}
where the function $\lambda(r)$ is given by
\begin{equation}
\lambda(r) = \sqrt{{D_{ABC}\tilde H^A(r)\tilde H^B(r)\tilde 
H^C(r)\over H_0(r)}} .
\end{equation}
Then
the four-dimensional metric of the extremal
black-hole  is given by 
\begin{equation}
e^{-2U(r)}=2\sqrt{H_0(r)\,D_{ABC}\,\tilde H^A(r)\,\tilde H^B(r)\,\tilde 
H^C(r)} ,
\label{IIAmetric}
\end{equation}
and the scalar fields have the following radius dependence:
\begin{equation}
z^A(r)=i\tilde H^A(r)\sqrt{{H_0(r)\over D_{ABC}\tilde H^A(r)\tilde H^B(r)
\tilde H^C(r)}}.
\end{equation}
The scalars at the horizon are determined as
\begin{equation}
z^A_{\rm hor}={Y^A_{\rm hor}\over Y^0_{\rm hor}}\,,\quad
Y^A_{\rm hor}={1\over 2} i p^A, \quad Y^0_{\rm hor}={1\over 2}
\sqrt{D\over q_0}\,,\quad D=D_{ABC}\,p^Ap^Bp^C\, .\label{IIAscal} 
\end{equation}
Finally, the corresponding macroscopic entropy takes the form 
\cite{BCDWKLM}
\begin{equation}
{\cal S}_{\rm BH}=2\pi\sqrt{q_0D}.\label{IIAentropy}
\end{equation}

\subsubsection{Topology change at the conifold point in type IIB 
Calabi-Yau compactications}

Now consider  type IIB compactifications on a Calabi-Yau space; hence
$N_V=h^{(2,1)}$.
The symplectic vector $V$  corresponds to the period intergrals
over the Calabi-Yau 3-cycles $\gamma_I$ and $\delta_I$:
\begin{equation}
 F_I = \int_{\gamma_I} \Omega \quad , \quad X^I = 
\int_{\delta^I} \Omega \ 
\end{equation}
The conifold transitions
occur at those points in the moduli space where
certain 3-cycles shrink to zero size and then blowing them up as two
cycles, changing in this way the Hodge numbers. 
In the following we will discuss the most simple situation with  periods
$X^0$ and $X^1$ (together with $F_0$ and $F_1$), where $X^1$ vanishes
at the conifold point and $X^0$ remains finite. 
Near this point
the prepotential can be expanded as
\begin{equation}
F  = -i \, (Y^0)^2 \left( c + {1 \over 4 \pi}  \biggl({Y^1\over Y^0}\biggr)^2 
\log i {Y^1\over Y^0}\,  + (\mbox{analytic terms})  \right) \ .
\end{equation}
($c= \chi \zeta(3) / 2 (2\pi)^3$).

Let us look on the 
structure of the space-time solution near the conifold point \cite{BehLu}.
For simplification, we will consider again axion-free black holes with
non-vanishing charges $q_0$ and $p^1$.
As solution of the stabilization equations (\ref{stable}) we get
\begin{equation} 
Y^0(r) = {\lambda (r)\over 2} \qquad , \qquad Y^1(r) = 
i {\tilde H^1 (r)\over 2}
\end{equation}
where now the function $\lambda(r)$ is given by
\begin{equation}
\lambda(r) = {H_0(r) \over 2 c } -  \, {( \tilde H^1(r))^2\over 4\pi \, 
H_0(r)} + 
{\cal O}((\tilde H^1(r))^4) \ .
\end{equation}
Keeping only the first correction, we obtain for
the function $e^{-2U}$ in the metric   
\begin{equation}
e^{-2U(r)} =
{H_0^2(r) \over 4c} + { (\tilde H^1(r))^2 \over 4\pi}  \, 
\log {2c\, {\tilde H^1(r)}\over H_0(r)} \ . 
\end{equation}
Hence, we obtained a non-singular metric, also at the points where
3-cycles vanish (${\tilde H^1}=0$). 

As a special case consider the black hole solution with
only non-vanishing charge $p^1$, i.e. $q_0=0$. Furthermore we also set
$\tilde h^1=0$. Then we obtain a {\it massless}
black hole with metric and scalar field $z^1$ as follows:
\begin{equation}
e^{-2U(r)} = 1 + {(p^1)^2 \over 8 \pi r^2} \log {c (p^1)^2 \over r^2 } \pm ..
\quad , \quad z^1 = i {\sqrt {c} \, p^1 \over r ( 1 \mp ..)} 
,
\end{equation}
where we used $h_0 = 2 \sqrt{c}$,  in order to have an asymptotic
Minkowski space and $\pm\cdots $ indicate higher powers in $1/r$.
This black hole solutions corresponds to a single D 3-brane which is
wrapped around the shrinking 3-cycle.
It is the dual to the electric  massless BPS state discussed by
Strominger.
The main property of this massless solution is that it
carries only one charge and has a shrinking internal
3-cycle at spatial infinity ($z^1 \to 0$ for $r \to \infty$).
Hence the Calabi-Yau space degenerates at $r=\infty$ 
where the topology change can
take place.

\subsection{Higher curvature corrections and microscopic $N=2$ entropy}

The $N=2$ black holes together with their entropies which were 
considered so far, appeared as solutions of the equations of 
motion of 
$N=2$ Maxwell-Einstein supergravity action, where the
bosonic part of the action contains terms with
at most two space-time derivates (i.e., the Einstein action, 
gauge kinetic terms and the scalar non-linear $\sigma$-model).
However,
the $N=2$ effective action 
of strings and M-theory contains in addition an infinite number 
of higher-derivative terms involving higher-order products of 
the Riemann tensor and the vector field strengths.
such as nonminimal gravitational couplings 
$R^{2}$. 
A particularly interesting subset of these couplings in $N=2$ 
supergravity can be again 
described by a holomophic function $F(X,W^2)
=\sum_{g=0}^\infty F^{(g)}(X^I)W^{2g}$, as discussed in section (3.4).
Using the superconformal calculus, 
$N=2$ black-hole solutions  for higher
order $N=2$ supergravity based on the holomorphic function $F(X,W^2)$
were studied in ref.\cite{msw,higher}.
As a result of these investigations,
the 
Bekenstein-Hawking entropy of $N=2$
black holes will receive corrections due to the higher derivative terms
in the effective $N=2$ supergravity action. 
In case of type IIA Calabi-Yau compactifications the higher order 
entropies will depend on additional topological Calabi-Yau data,
such as the second Chern numbers or the elliptic instanton numbers, since
the functions $F^{(g)}$ depend on these data (for $F^{(1)}$ see 
eq.(\ref{f1typeII})).
These corrections are essential to match the macroscopic
Bekenstein-Hawking entropy with the microscopic black hole
entropy, which comes from counting the corresponding D-brane
degrees of freedom. In fact, for the large radius limit of type IIA,
Calabi-Yau 
compactifications, the microscopic entropy was recently computed 
\cite{msw}:
\begin{equation}
{\cal S}_{\rm micro}=2\pi\sqrt{q_0
D\Big(1+{c_{2A}\,p^A\over 6D}\Big)}\, .\label{malmicro}
\end{equation}
The $c_{2A}$ are the second Chern class numbers of the Calabi-Yau 3-fold.
As shown in \cite{msw,higher}, expanding this expression to lowest order
in $c_{2A}$, 
\begin{equation}
{\cal S}_{\rm micro}=2\pi\sqrt{q_0 D}+
2 \pi 
{1\over 12}c_{2A}\,p^A
\sqrt{{q_0\over D}}  + \cdots \ ,\label{entr1msw}
\end{equation}
the correction  
agrees
with a correction to the effective action involving $R^2$ type terms
with a coefficient function $F^{(1)}(z^A)=-{1\over 24}c_{2A}z^A$
(see eq.(\ref{f1typeII})).
In order to match the full microscopic $N=2$ entropy from the effective action,
more work is still required.

\section{Summary}

In this article we have reviewed the string-string duality between string vacua
with $N=2$ supersymmetry or, respectively, with $N=1$ supersymmetry in six
dimensions. All the results presented here support the hypothesis of
the $N=2$ string-string duality symmetry.
Beyond checking the string-string duality for a large class of heterotic/type
II vacua, new insights into non-perturabtive phenomena in 
compactified string theory
are gained. 
Many of these results have a very beautiful description in terms
of Calabi-Yau geometry. It appears that there there is a huge web 
of $N=2$ string theories, continuously connecting
 many,
perhaps all, $N=2$ vacua in four dimensions.
Especially, in the study of the moduli spaces of type II
strings compactified on Calabi-Yau three-folds it became clear that the
topologies of the Calabi-Yau spaces can be continuously deformed into
each other.
  The transitions
among different vacua are made possible by BPS states which become massless
at some special loci in the moduli spaces where the transitions take place.
In the geometric type II Calabi-Yau description these BPS states originate
from wrapping the ten-dimensional D p-branes around internal cycles of
the Calabi-Yau space. 
Seen from the four-dimensional point of view, the BPS states correspond
to supersymmetric charged black hole solutions of the effective $N=2$ 
supergravity action.

Of even greater phenomenological interest than $N=2$ string vacua
is the investigation
of string duality symmetries in four-dimensional string vacua with
$N=1$ space-time supersymmetry. One hopes in this way to get
important nonperturbative information, like the computation of
nonperturbative $N=1$ superpotentials \cite{wsup} 
and supersymmetry breaking, or the
stringy reformulation of many effects in $N=1$ field theory.
In addition, it is a very important question, how 
transitions among $N=1$ string vacua,
possibly with a different number of chiral multiplets, take place 
\cite{n1trans}.
String dual pairs with $N=1$ supersymmetry in four dimensions 
are
provided by comparing heterotic string vacua on  Calabi-Yau three-folds,
supplemented by the specification of a particular choice of the
heterotic gauge bundle, 
with particular four-dimensional type IIB
superstring vacua, which can be formulated as $F$ theory
compactifications  on elliptic Calabi-Yau four-folds 
\cite{V,fourfolds,n4pairs}.
In \cite{n4pairs} we have constructed several $N=1$ dual string pairs with
identical massless spectra. Moreover, one can show that for a particular
class of dual $N=1$ string vacua the $N=1$ superpotentials agree.
These superpotentials are given by specific modular functions, very
similar like the $N=2$ prepotentials we discussed here.
It would be interesting to gain further insights into some common features
of $N=2$ prepotentials and $N=1$ superpotentials from string theory.

\vskip0.5cm
\noindent {\bf Acknowledgements}
\vskip0.2cm
\noindent I like to thank my collaborators 
for numerous discussion on the material, presented
here.

\end{document}